\documentclass[fleqn,usenatbib]{mnras}

\usepackage[T1]{fontenc}
\usepackage{ae,aecompl}
\usepackage{booktabs}
\usepackage{array}
\usepackage{cancel}

\makeatletter
\@ifundefined{abovecaptionskip}{\newskip\abovecaptionskip}{}
\@ifundefined{belowcaptionskip}{\newskip\belowcaptionskip}{}
\makeatother

\usepackage{graphicx}	
\usepackage{amsmath}	
\usepackage{amssymb}	
\usepackage{soul}
\usepackage[normalem]{ulem}
\usepackage[dvipsnames]{xcolor}
\usepackage[normalem]{ulem}
\usepackage{cancel}
\usepackage{lineno}
\usepackage{float}
\usepackage{booktabs}
\usepackage{multirow}
\usepackage{stfloats}
\usepackage{placeins}

\makeatletter
\renewenvironment{figure*}[1][tbp]
  {%
    \st@rredfloattrue
    \let\@makecaption=\SFB@makefigurecaption
    \@dblfloat{figure}[#1]%
    \centering
    \fstyle@figure
  }
  {%
    \end@dblfloat
    \st@rredfloatfalse
  }
\makeatother
\restylefloat{table}
\definecolor{purple}{rgb}{0.5,0,0.87}

\definecolor{royalblue}{rgb}{0.0, 0.14, 0.4}

\linenumbers

\usepackage{newtxtext,newtxmath}
\usepackage{orcidlink}

\title[Intracluster light in the infall region]{Intracluster light in the infall region of \\ simulated protoclusters of galaxies \\[-0.7em]}

\author[S. V. Werner et al.]{S. V. Werner$^{1,2}$\thanks{E-mail: stephanevazwerner@gmail.com} \orcidlink{0000-0002-7186-7889},  M. Jauzac$^{3,1,2,4,5}$ \orcidlink{0000-0002-7186-7889}, M. Montes$^{6}$ \orcidlink{0000-0001-7847-0393}, D. Rennehan$^{7}$ \orcidlink{0000-0002-1619-8555},
\newauthor
G. Mahler$^{8}$ \orcidlink{0000-0003-3266-2001},
A. Niemiec$^{9}$ \orcidlink{0000-0003-3791-2647},  Y. Jim\'enez-Teja$^{10,11}$ \orcidlink{0000-0002-6090-2853} 
\newauthor \vspace{-0.2cm}  \\
$^{1}$ Centre for Extragalactic Astronomy, Department of Physics, Durham University, South Road, Durham DH1 3LE, UK\\
$^{2}$ Institute for Computational Cosmology, Department of Physics, Durham University, South Road, Durham DH1 3LE, UK\\
$^{3}$ Univ Toulouse, CNES, CNRS, IRAP, Toulouse, France\\
$^{4}$ Astrophysics Research Centre, University of KwaZulu-Natal, Westville Campus, Durban 4041, South Africa\\
$^{5}$ School of Mathematics, Statistics and Computer Science, University of KwaZulu-Natal, Westville Campus, Durban 4041, South Africa\\
$^{6}$ Institute of Space Sciences (ICE, CSIC), Campus UAB, Carrer de Can Magrans, s/n, 08193 Barcelona, Spain \\
$^{7}$ Center for Computational Astrophysics,
Flatiron Institute, 162 5th Ave, New York, NY 10010, USA \\
$^{8}$ STAR Institute, Quartier Agora - All\'ee du six Ao\^ut, 19c B-4000 Li\`ege, Belgium\\
$^{9}$ Univ. Grenoble Alpes, CNRS, Grenoble INP, LPSC-IN2P3, 38000 Grenoble, France\\
$^{10}$ Instituto de Astrof\'isica de Andaluc\'ia--CSIC, Glorieta de la Astronom\'ia s/n, E--18008 Granada, Spain \\
$^{11}$ Observat\'orio Nacional, Rua General Jos\'e Cristino, 77 - Bairro Imperial de S\~ao Crist\'ov\~ao, Rio de Janeiro, 20921-400, Brazil \\ \vspace{-0.9cm}}

\date{ Accepted XXX. Received YYY; in original form ZZZ \vspace{-0.5cm}}

\pubyear{2026}

\usepackage{indentfirst}
\begin{document}

\nolinenumbers
\label{firstpage}
\pagerange{\pageref{firstpage}--\pageref{lastpage}}
\maketitle

\begin{abstract}
Intracluster light (ICL) is a diffuse stellar component in galaxy clusters, made of stars following the potential of the cluster. Despite its potential to unveil the accretion history of clusters, the ICL in protoclusters remains poorly understood in simulations and observations. We analyse protoclusters in TNG-Cluster and Manhattan Suite simulations to investigate when ICL forms, its host progenitors, and environments where it forms. We find a considerable amount of ICL in the infall region of protoclusters at 2<z<4, ranging from $\rm 0.1\times10^{11} \! < \! M_{*}/M_{\odot} \! < \! 3\times10^{11}$. ICL star formation in protoclusters peaks at $3<z<4$, with stars formed in the main halo forming earlier than those born in the infall regions. We find that $74.2_{-1.7}^{+1.7}\%$ of ICL stars in TNG-Cluster and $69.2_{-3.7}^{+3.6}\%$ in Manhattan Suite were stripped from galaxies within the main halo. However, most ICL stars in protoclusters at $\rm z=2$ formed in the infall region, accounting for $57.6_{-1.6}^{+1.7}\%$ and $55.6_{-2.3}^{+2.3}\%$ in TNG-Cluster and Manhattan Suite, respectively. By the time the ICL stars are stripped, their host galaxies are $3$--$4$ times more massive than the galaxies in which the stars originally formed, implying growth of the progenitor galaxies and/or accretion of the stars onto more massive galaxies prior to stripping. The ICL stellar masses measured in simulations agree with observations at z$\leq$2. Our findings highlight the importance of investigating the diffuse light in the outer regions of protoclusters to constrain the formation and assembly of the ICL and galaxy clusters.

\end{abstract}




\section{Introduction} \label{sec:intro} 


There is a faint and diffuse stellar component that permeates galaxy clusters, known as the intracluster light (ICL). It was first proposed by Fritz Zwicky in 1937 while he was studying the missing mass problem in galaxy clusters \citep{Zwicky_1937}. Although the ICL could not fully explain the missing mass (now attributed to dark matter), Zwicky observed ICL years later in the centre of the Coma cluster \citep{Zwicky_1951}. Over the years, studies have determined that this faint light is composed of stars that are not connected to galaxies, but are instead bound to the galaxy cluster potential \citep{Miller1983, Merritt_1984, Theus1997, Montes_2019, Montes_2022, Contini_2013, Contini_2014, Contini_2018}. Although this diffuse component is hard to detect because its low surface brightness ($\mathrm{\mu_{V}\!\gtrsim\! 27 mag/arcsec^{2}}$) requires deep imaging and its extent over hundreds of kiloparsecs demands wide-field observations, advances in telescope instrumentation and detection techniques are enabling increasingly detailed studies of how these stars assemble in galaxy clusters \citep{Mihos_2019}. The ICL is connected to the evolution of the central cluster galaxy and represents a powerful way of testing galaxy evolution models \citep{Burke_2015, Demaio2020}. It also gives hints on the accretion history of clusters \citep{Merritt_1984, Conroy_2007, Contini_2021_review, Kimmig_2025}.

Recent studies have shown that most of these stars were stripped from galaxies during interactions between them, and several physical mechanisms were proposed to explain how these stars were stripped from galaxies, including mergers \citep[e.g.][]{Conroy_2007, Murante_2007, Joo_2025}, tidal interactions \citep[e.g.][]{Rudick_2009, Contini_2014, Contini_2018, Contini_2019, Martin_2024}, and dwarf disruptions \citep[e.g.][]{Purcell_2007}.
Some observations suggest that \textit{in-situ} formation is responsible for less than 1\% of the ICL mass \citep[e.g.][]{Melnick_2012}, while others using simulations argue that this fraction can reach 28$\%$ \citep[e.g.][]{Ahvazi_2024}. However, observations have so far provided only limited constraints on the contribution of \textit{in-situ} star formation to the ICL, despite detections of localized star formation in intracluster environments \citep[e.g.][]{Gerhard_2002, Sun_2007, Barfety_2022}.

Another relevant process for ICL assembly is pre-processing. Like its galaxy counterpart \citep{Lewis2002, Gomez2003, Patel2011, Oemler2013, Haines2015, Bianconi2018, Just2019, Werner_2021}, ICL pre-processing is the building of ICL in infalling groups of galaxies that will end up as ICL in the cluster after merging \citep{Mihos_2004, Contini_2024}. However, the relevance of this channel for ICL formation is still debated. \cite{Mihos_2004} discussed that groups allow for slow encounters between galaxies in clusters which can cause pre-processing of the ICL. More recently, using Euclid data, \cite{Ellien_2025} found that groups account for $\sim$30\% of the final ICL within 300 kpc from the cluster center in Abell 2390, while using simulations \citet{Contini_2024} found 20\%. \cite{Jimenez_2025} found that galaxy groups infalling into the Coma cluster contained red intragroup light (IGL), indicating that their diffuse light already underwent at least some amount of pre-processing. Considering that ICL is in general mostly concentrated around the Brightest Cluster Galaxy (BCG), some works consider the system BCG+ICL as a single component, given the difficulty in defining the limit between the BCG and the ICL. \cite{Jeon_2026} used the \textit{NEWCLUSTER} simulation and found that $12.2\%$ of the BCG+ICL component within 1.5$\times\mathrm{R_{200}}$ was pre-processed at $z\!=\!0.79$, while $32.5\%$ formed \textit{in-situ}.

There are many analyses that measure ICL fractions using observations and simulations \citep{Montes_2022, Contreras_Santos_2024, Brough_2024}, but most focus on the evolution at low redshifts ($z\!\lesssim\!1$). More recently, despite the significant observational challenges, some works have shown evidence for ICL in high redshift clusters at $1\!<\!z\!<\!2$ \citep{Demaio2020,Joo_2023, Ko_2018, Burke_2012}. This includes the measurement of ICL in a galaxy group at $z\!\sim\!1.85$ by \citet{Coogan_2023}, as well as measurements by \citet{Werner_2023} in the cores of two $z\!\sim\!2$ protoclusters (XLSSC 122 and CARLA J1018). The former was recently targeted with JWST data by \citet{Joo_2026} to estimate its outer surface brightness profiles and fractions.

Because of the observational challenges, it has proved challenging to compile an observational picture of ICL formation and evolution at earlier times due to the lack of a large sample of protoclusters at $z\!>\!2$. However, focused cosmological simulations can provide insights into possible mechanisms involved in the origin and evolution of the ICL in galaxy clusters. We can obtain predictions of how ICL stars form and assemble in clusters in regimes that are still not accessible to observations such as at high redshift. In this work, we aim to use simulations to answer the following questions:

\begin{itemize}
    \item[--] When and where did the stars in the ICL in protoclusters form? 
    \item[--] What is the typical stellar mass of a galaxy that hosted ICL in the past? Does the galaxy in which an ICL stellar particle formed have the same properties as the galaxy from which it was later stripped?
    \item[--] Is there ICL in the infall region of protoclusters? If so, what is the ICL stellar mass in this region?
    \item[--] Is there a correlation between the distribution of stars in protoclusters and where they form?
\end{itemize}

Here, we use the Manhattan Suite \citep{Rennehan_2024} and the TNG-Cluster simulations \citep{nelson2021, nelson2024} to study the formation and assembly of ICL in protoclusters, and try to answer the questions listed before. In Section \ref{simulations}, we describe the zoom-in protoclusters simulations used in this work. In Section \ref{sec:methods}, we describe how ICL particles are selected and define the different environments. We also provide a description of how particles are being tracked, and how we measure their distances to cluster centers. Our results on the location and timescale under which the ICL was formed in Section \ref{sec:results}. A discussion and comparison with previous works are presented in Section \ref{sec:discussion}. In Section \ref{sec:conclusions}, we summarize our main findings and conclusions.

\vspace{-0.3cm}
\section{Simulations}

\label{simulations}

\begin{figure}
\centering 
\hspace{-1.2cm}
\includegraphics[width=0.9\columnwidth]{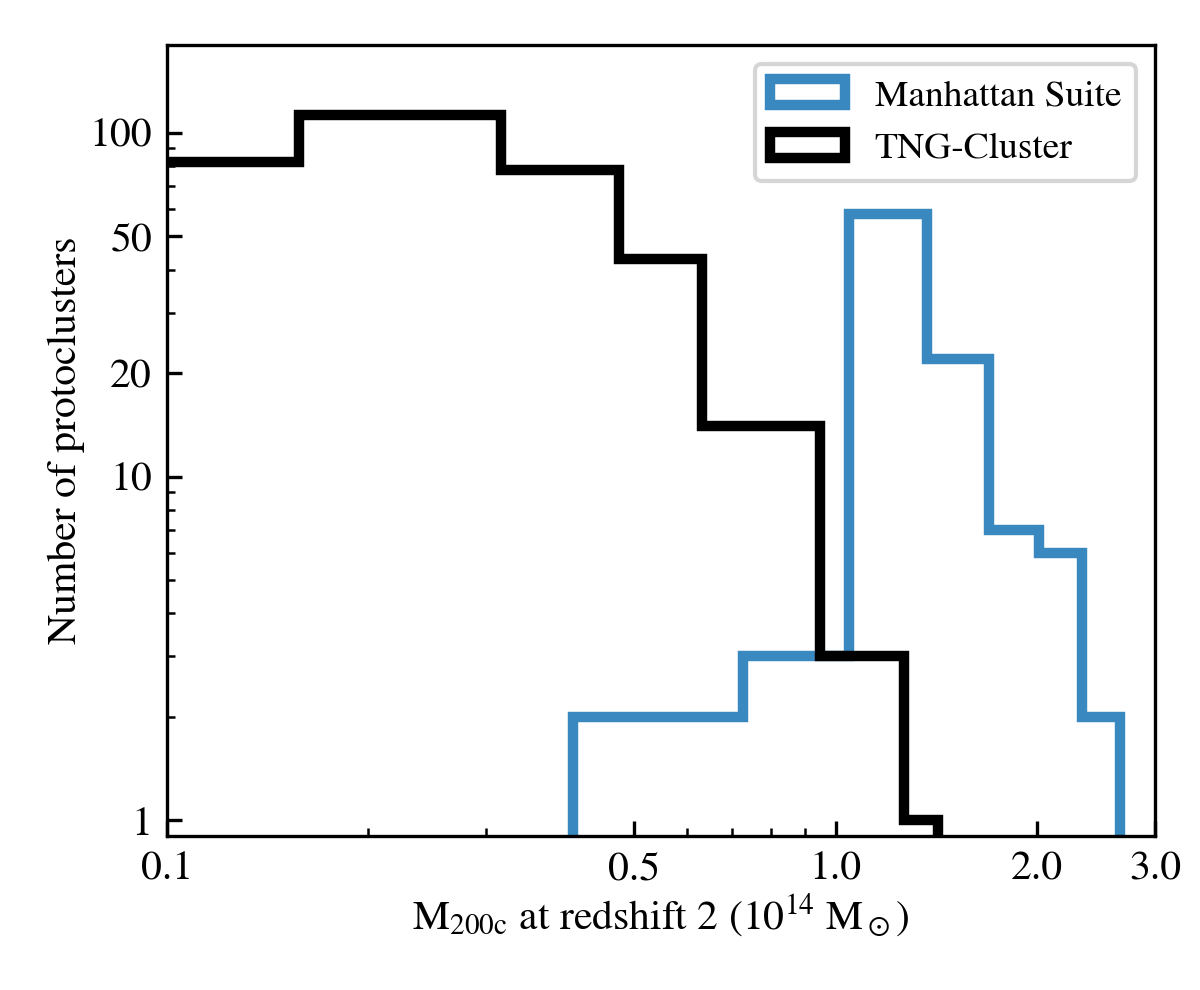}
\vspace{-0.3cm}
\caption{Mass distribution of protoclusters at z$\sim$2 for the Manhattan Suite (blue) and TNG-Cluster (black). $\rm M_{200c}$ is the mass in which the density is 200 times the critical density of the Universe at $z=2$. The Manhattan Suite sample is generally more massive than the TNG-Cluster sample. The difference between the two distributions is due to a selection effect, i.e., the Manhattan Suite protoclusters were selected from a larger volume $(1.5\,\mathrm{cGpc})^3$ and as the most massive protoclusters at $z=2$. On the other hand, TNG-Cluster protoclusters are the most massive at redshift 0 and from a smaller volume $(1.0\,\mathrm{cGpc})^3$. This makes the two samples complementary.}
\label{massdist}
\end{figure}

In this work, we use two simulation suites: the Manhattan Suite \citep{Rennehan_2024} and TNG-Cluster \citep{nelson2021, nelson2024}. Fig.~\ref{massdist} shows that these simulations have different selection biases in their mass distributions at $z=2$. The Manhattan Suite protoclusters were selected from a larger volume $(1.5\,\mathrm{cGpc})^3$ and as the most massive protoclusters at $z=2$. On the other hand, TNG-Cluster protoclusters are the most massive at redshift 0 and from a smaller volume $(1.0\,\mathrm{cGpc})^3$. Besides that, TNG-Cluster has additional information about clusters at redshifts lower than 2. These two characteristics make TNG-Cluster and Manhattan complementary samples. It is also important to note that this is the first analysis of the ICL using both simulations. More details are given below.

\vspace{-0.2cm}
\subsection{The Manhattan Suite} 
\label{sec:TMS}

\begin{figure*}
\centering
\includegraphics[width=1.6\columnwidth]{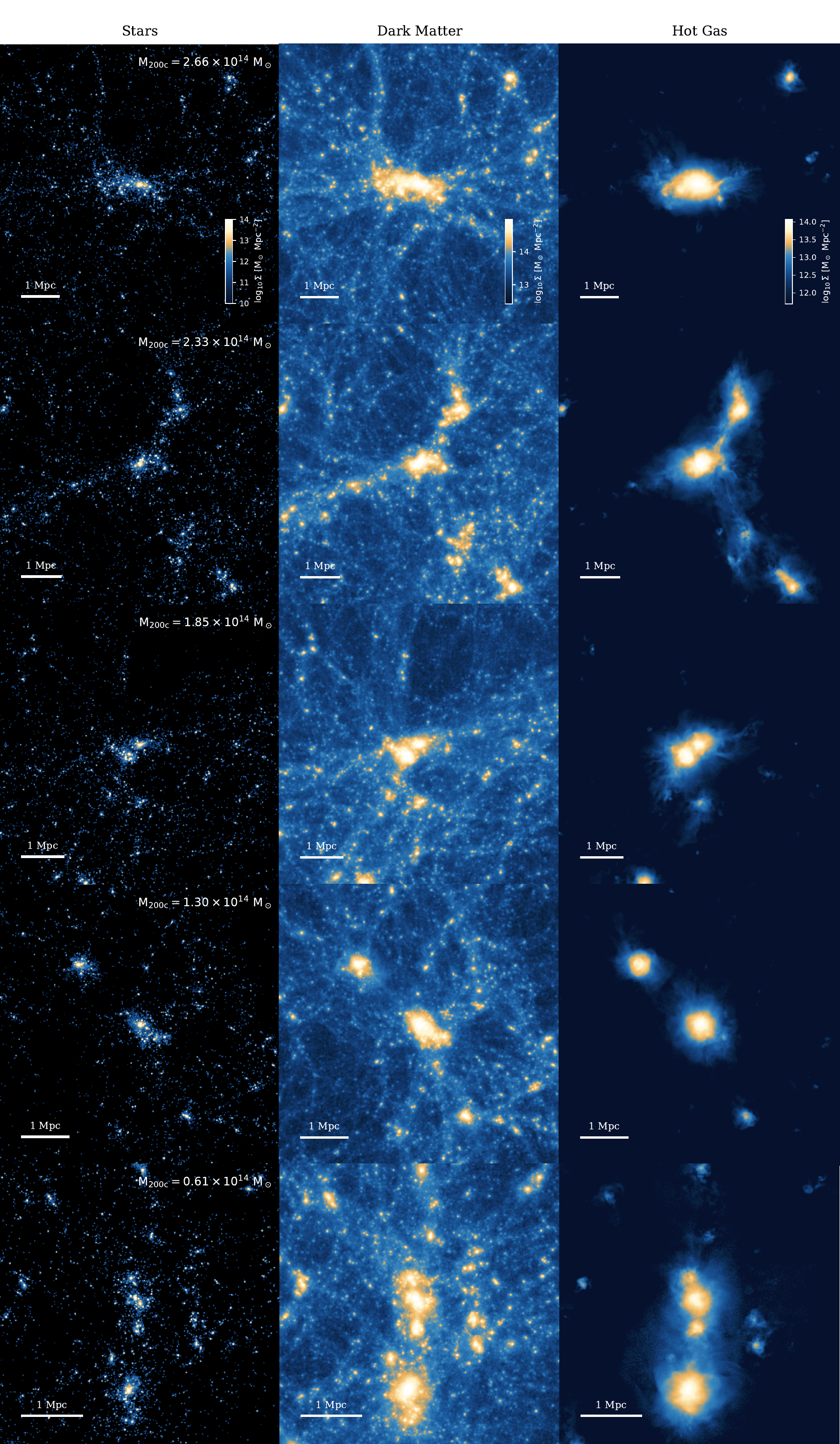}
\caption{The spatial distribution of stars, dark matter and hot gas in protoclusters in the Manhattan Suite simulation at z$\sim$2. The hot gas is selected according to its temperature ($\mathrm{T>10^{7} K}$). The $\rm M_{200c}$ at $z=2$ for each protocluster are shown in the left panels. We also add a scale bar of 1 Mpc for comparison. Most protoclusters are dynamically young and are accreting structures, such as massive galaxies and groups, and/or going through major merger events.} 
\label{manhattanproto}
\end{figure*}

\textit{The Manhattan Suite} is a set of $100$ zoom-in simulations of protoclusters that is an intentionally biased sample to the most massive galaxy protoclusters at $z\!=\!2$ \citep{Rennehan_2024}. In particular, they selected the most massive haloes at $z \! = \! 2$ from a $(1.5 \,\mathrm{cGpc})^3$ N-body pre-flight simulation and resimulated them at significantly higher resolution. The mass resolutions of gas, star and dark matter particles in Manhattan are $\rm m_{gas} \! \sim \! 3.5\times10^{7}\,M_{\odot}$, $\rm m_{star} \! \sim \! 3.5\times10^{7}\,M_{\odot}$ and $\rm m_{DM} \! = \! 1.9\times10^{8}\,M_{\odot}$ respectively. At $z = 2$, all of the resimulated haloes have virial masses $\mathrm{M_{vir} \! >  \! 10^{14}}\,\mathrm{M}_\mathrm{\odot}$. Fig.~\ref{manhattanproto} shows projected visualizations of the stellar, dark matter, and hot gas components in protoclusters of the Manhattan Suite, masses are given in the top right of the left panels.

\begin{figure}
\centering
\includegraphics[width=1\columnwidth]{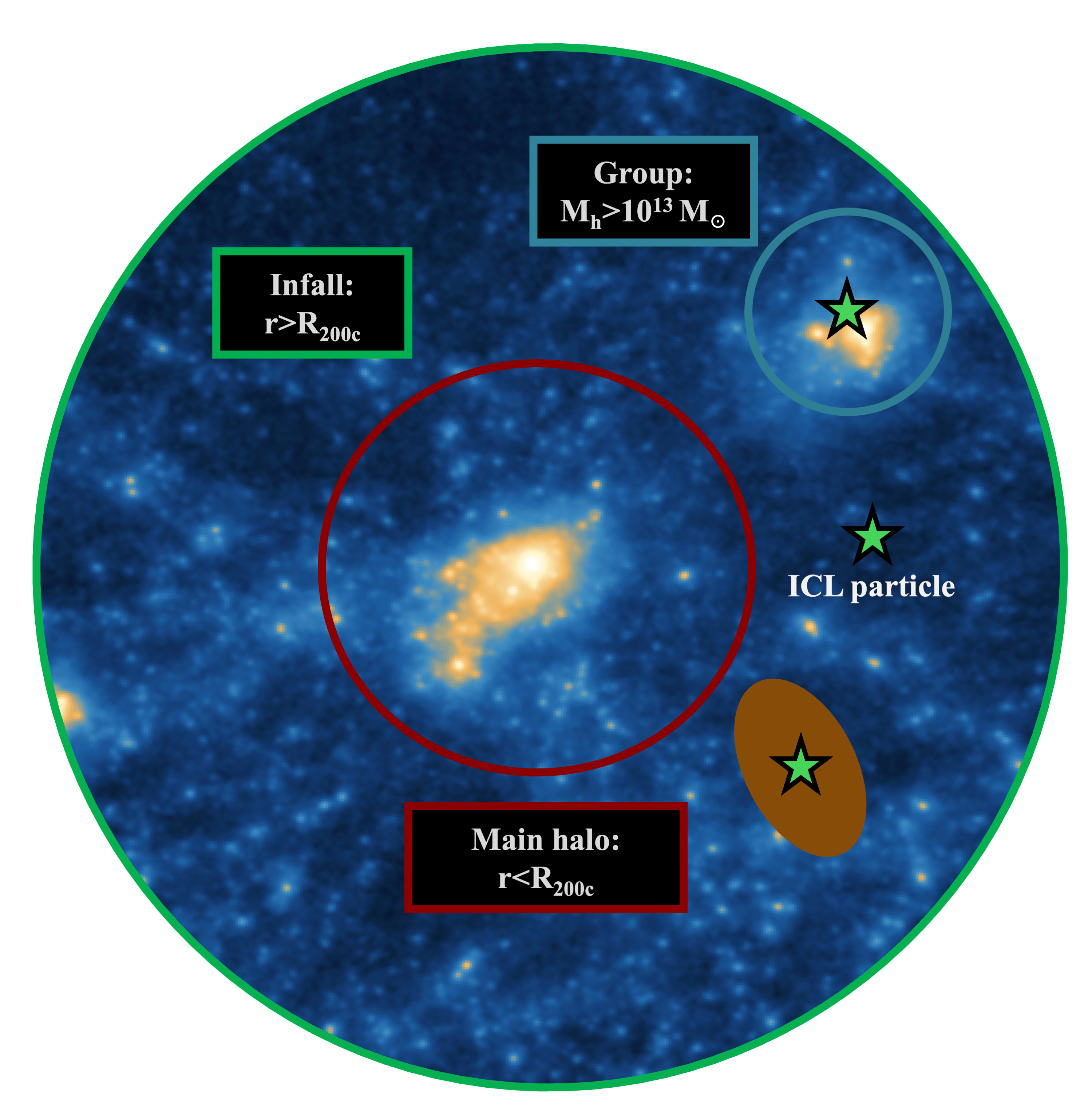}
\vspace{-0.5cm}
\caption{The scheme illustrates our environment classification. We show the environment definition for main halo, infall and group. The green stars represent ICL stellar particles, while the orange ellipse represents a protocluster galaxy. The stellar particle in the infall can be formed inside or outside a group and/or in or outside a galaxy. Note that the sizes are not to scale. }
\label{defs}
\end{figure}

\begin{figure*}
\centering
\makebox[\textwidth][c]{%
    \hspace*{2cm}%
    \includegraphics[width=2\columnwidth]{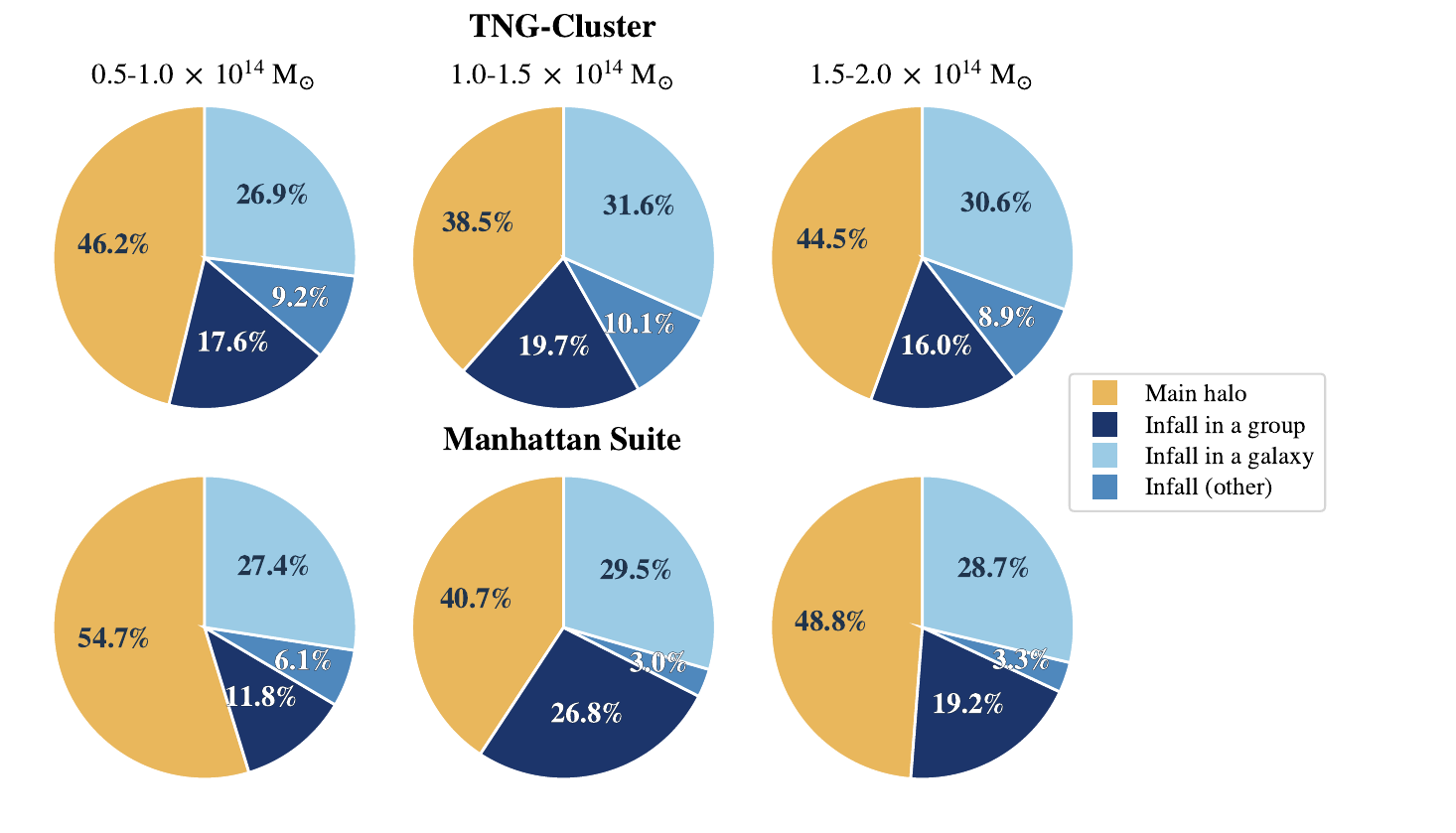}%
}
\vspace{-0.7cm}
\caption{Distribution of ICL stellar particles depending on where they formed in the TNG-Cluster (top) and Manhattan Suite (bottom). Once the formation redshift of each stellar particle is selected, we classify its location as described in Section \ref{def_envi}. The pie charts show where the ICL mass was first formed and if they were ever associated with a group. The total mass of each environment is divided by the total ICL mass at z$\sim$2. The orange colour means the ICL stellar particle was created in the main halo, while the blue colours mean the ICL particle was created in the infall region. Then we have the subclassifications for those born in the infall region: (i) it can be part of a group at some point (dark blue); (ii) never been in a group and formed inside a galaxy (light blue) and (iii) never been in a group and formed outside a galaxy (medium blue). Some stars may have formed within the group, while others may have joined it later. The pie charts correspond to different protocluster mass ranges in $\rm M_{200c}$ at $z=2$, in which the left one has the lowest mass range and the right one the highest mass range. The values shown in the pie charts correspond to the mean values of all protoclusters in each mass bin. Note that these are mass weighted estimates considering the mass of all stellar particles. We have 20 (3), 20 (20), 6 (17) protoclusters in each mass bin for TNG-Cluster (Manhattan). Most ICL particles were born in the infall region.}
\label{piechart}
\end{figure*}

\begin{figure*}
\includegraphics[width=2.0\columnwidth]{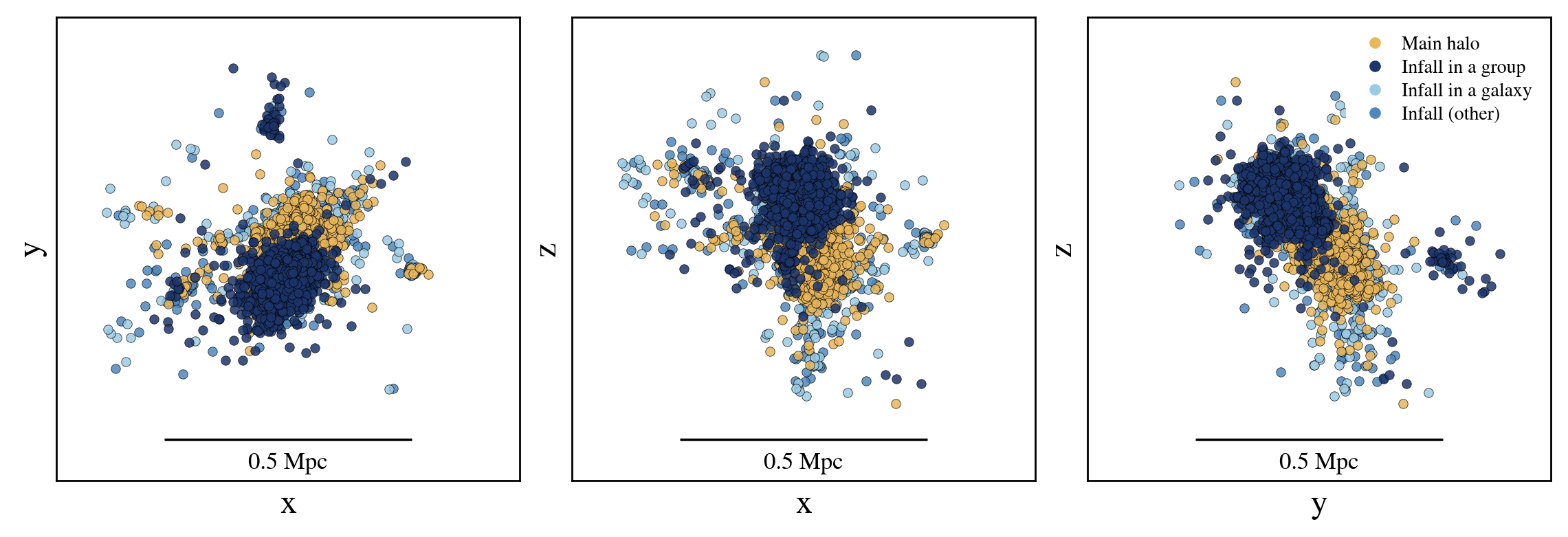}
\vspace{-0.3cm}
\caption{Projected distribution of ICL particles within $\rm R_{200c}$ at z$\sim$2 for a galaxy protocluster in TNG-Cluster ($\rm M_{200c}\! \sim \! 7.5\times10^{13}M_{\odot}$). Colours show different environments where these particles were created in the simulation. Classifications and colours are the same as for Fig.~\ref{piechart}. We can clearly see that a fraction of ICL stars were located in groups and later merged into the protocluster main halo.}
\label{infall_def}
\end{figure*}

The Manhattan Suite uses the \textit{Simba} galaxy formation model \citep{Dave2019}.  \textit{Simba} includes models for cooling, star formation, stellar feedback, and black hole evolution and stellar feedback built into \textsc{GIZMO} \citep{Hopkins2015a}. The simulations were run using the mesh-free finite mass (MFM) hydrodynamics method \citep{Lanson2008a, Lanson2008b, Gaburov2011}, a publicly available hydrodynamics+gravity solver. We point the reader to \cite{Hopkins2015a, Hopkins2016b}, \cite{Dave2019}, and \cite{Rennehan_2024} for a more complete description. Manhattan provides 63 snapshots between redshift 12.7 and 2.03, with a median time resolution of 34.16 Myr.

The structure catalogues in the Manhattan Suite were computed using the \textsc{caesar}\footnote{\url{https://caesar.readthedocs.io/}} package, a friends-of-friends (FoF) galaxy and halo finder.  In particular, \textsc{caesar} finds galaxies by searching for groups of stars and cold gas ($n_\mathrm{H} \! > \! 0.13 \, \mathrm{cm}^{-3}$, $T \! < \! 10^{5}\,\mathrm{K}$) and associates them with the nearest FoF dark matter halo minimum potential center. 


Recall that the Manhattan Suite sample was selected as the $100$ most massive structures at $z \! = \! 2$ in a large sample volume, $(1.5\,\mathrm{cGpc})^3$. This has the advantage of encoding the observational bias one would expect at $z \! \sim \! 2$, in which we observe the brightest, most massive galaxy clusters. The fundamental assumption is that the most massive galaxy clusters at $z \! = \! 2$ would have the brightest and most massive protoclusters at $z \!  > \! 2$. 

The cosmology of the simulation is consistent with \cite{Ade2016}, having $\rm \Omega_{m,0} \! = \! 0.308$, $\rm \Omega_{\Lambda,0} \! = \!  0.692$, $\rm h \! = \! 0.6781$, $\rm \Omega_\mathrm{b,0} \! = \! 0.0484$, $\rm n_\mathrm{s} \! = \! 0.9677$, and $\rm \sigma_{8} \! = \! 0.8149$.


\vspace{-0.2cm}
\subsection{TNG-Cluster} 
\label{sec:TNG}

TNG-Cluster is a cosmological magnetohydrodynamical simulation of galaxy formation that resimulates $352$ galaxy clusters \citep{nelson2021, nelson2024}. The advantage of this sample is that it provides clusters with $\mathrm{M_{halo}} \! > \! 10^{14}\,\mathrm{M_{\odot}}$ up to $z \! = \! 0$ in a $(1\,\mathrm{cGpc})^{3}$ volume. The mass resolutions of baryon and dark matter particles in TNG-Cluster are $\rm m_{baryon} \! \sim \! 1.2\times10^{7}\,M_{\odot}$ and $\rm m_{DM} \! = \! 6.1\times10^{7}\,M_{\odot}$ respectively.At $z \! = \! 2$, TNG-Cluster has 352 protoclusters in total, including 209 with $\mathrm{M_{200c}} < 0.5 \times 10^{14}\,\mathrm{M_{\odot}}$, $112$ protoclusters with $\mathrm{M_{200c}} \! \sim \! 0.5-1.0 \times 10^{14}\,\mathrm{M_{\odot}}$, 25 with $\mathrm{M_{200c}} \! \sim \! 1.0-1.5 \times 10^{14}\,\mathrm{M_{\odot}}$, 6 with $\mathrm{M_{200c}} \! \sim \! 1.5-2.0 \times 10^{14}\,\mathrm{M_{\odot}}$. TNG-Cluster provides 100 snapshots between redshifts 20.05 to 0, with mean time resolution of 146 Myr.

TNG-Cluster was built so we can deepen our understanding of the formation of BCGs and satellite galaxies, and the assembly of protoclusters of galaxies. Several works have shown that TNG-Cluster can give us valuable insights to understand the kinematics of the hot intracluster medium \citep{Ayromlou_2024}, cluster mergers \citep{Lee_2024}, and AGN feedback in galaxy clusters \citep{nelson2024}. Compared to Manhattan, it has the advantage of having information between $\mathrm{0 \! < \! z \! < \! 2}$, while Manhattan only provides information at $\mathrm{z>2}$, so we can compare our ICL measurements with previous observational works in the literature. More information about TNG-Cluster can be found in the TNG-Cluster website\footnote{\url{https://www.tng-project.org/cluster/}}.

The cosmology of the simulation is consistent with \cite{Planck_2016}, having $\rm \Omega_{m,0} \! = \!  0.3089$, $\rm \Omega_{\Lambda,0} \! = \! 0.6911$, $ \rm h \! = \! 0.6774$, $\rm \Omega_\mathrm{b,0} \! = \! 0.0486$, $\rm n_\mathrm{s} \!  = \! 0.9667$, and $\rm \sigma_{8} \! = \! 0.8159$. Although the two simulations use slightly different cosmological parameters, this difference is too small to affect any of the results presented in this paper.



%

\vspace{-0.4cm}
\section{Methodology} 
\label{sec:methods}

\begin{table*}
\centering

\begin{tabular}{lll p{0.50\textwidth}}
\toprule
Definition & Simulation & Catalogue & Operational definition \\
\midrule
Galaxy & Manhattan Suite & \textsc{caesar} &
Resolved galaxies with $\mathrm{M}_{\star} \! \geq \!  10^7\,\mathrm{M}_{\odot}$, identified from \textsc{caesar} stellar-particle membership lists. \\

\addlinespace[0.3cm]
Galaxy & TNG-Cluster & \textsc{subfind} &
Resolved stellar subhaloes with $\mathrm{M}_{\star} \! \geq \!  10^8.5\,\mathrm{M}_{\odot}$. Satellites are masked within
$\rm 2r_{\star,\mathrm{half}}$; the central galaxy is masked within
$\rm 3r_{\star,\mathrm{half}}$. \\

\addlinespace[0.3cm]
Group & Manhattan Suite & \textsc{caesar} halo catalogue &
External \textsc{caesar} halo with total mass $\mathrm{M}_{\mathrm{tot}} \!  \geq \!  10^{13}\,\mathrm{M}_{\odot}$, excluding the main halo. The group aperture is
$\rm R_{\mathrm{group}} \!  = \!  R_{200\mathrm{c},\mathrm{main}}(\mathrm{M}_{\mathrm{group}}/\mathrm{M}_{\mathrm{main}})^{1/3}$. \\

\addlinespace[0.3cm]
Group & TNG-Cluster & \textsc{subfind} subhalo catalogue &
Massive subhalo with total bound mass $\mathrm{M}_{\mathrm{sub}} \!  \geq \! 10^{13}\,\mathrm{M}_{\odot}$, excluding the main central subhalo. The group aperture is
$\rm R_{\mathrm{group}} \! = \! R_{200\mathrm{c},\mathrm{main}}(\mathrm{M}_{\mathrm{sub}}/\mathrm{M}_{200\mathrm{c},\mathrm{main}})^{1/3}$. \\

\bottomrule
\end{tabular}
\vspace{0.3cm}
\caption{Galaxy and group definitions adopted in this work. Galaxies are identified from the native catalogues of each simulation. Massive group/preprocessing structures are assigned an $\rm R_{200\mathrm{c}}$-equivalent aperture using the main-halo radius at the same snapshot. Manhattan provides memberships for each galaxy using \textsc{caesar}, while for TNG-Cluster we adopt a geometrical definition used in previous works.}
\label{table2}
\end{table*}

\subsection{Protocluster sample}

Protoclusters of galaxies can be defined in different ways such as: overdensities at z$>$2, progenitors of galaxy clusters at z$=$0 and/or massive structures at high redshifts. In this work, we select our protocluster sample by applying a mass cut at a fixed redshift, in which we only select structures with $\rm M_{200c}>0.5\times10^{14}\,M_{\odot}$ at $z=2$.

We select 20 protoclusters, when available, in each of three mass bins (0.5–1.0 $\rm \times 10^{14}\,M_{\odot}$, 1.0–1.5 $\rm \times 10^{14}\,M_{\odot}$, and 1.5–2.0 $\rm \times 10^{14}\,M_{\odot}$). This mass range is chosen because it contains protoclusters in both simulations; higher or lower mass ranges would not contain protoclusters from both simulations. Besides that, we select 20 clusters in each mass bin because the analysis is computationally expensive, so we wanted to minimize the run time while, at the same time, having statistically significant results. In the end, we have 20 (3), 20 (20), and 6 (17) protoclusters in TNG-Cluster (Manhattan) for the three mass bins, respectively. 

\subsection{Protocluster center definition}

We define the centre of the protocluster as the position of the most massive galaxy. Protoclusters are mostly dynamically young and are going through many mergers and accretion of groups, so the centering is an important definition since there might be a misalignment between the stellar distribution, the most massive galaxy position and the minimum potential. However, \cite{Werner_2023} showed that even in protoclusters that are still forming and contain merging structures, the ICL will mostly concentrate around the most massive galaxy of the merging systems, so this tight spatial correlation makes the most massive galaxy an excellent tracer for the surrounding ICL. Distances between ICL particles and the cluster centre were estimated using 3D distances, using the positions of the particles and the position of the BCG. 

\vspace{-0.3cm}
\subsection{Selection of ICL particles}

We define ICL particles as protocluster stellar particles that have not been assigned to a galaxy. For both simulations, we select target ICL particles at $z\simeq2$ within $\rm R_{200c}$ of the main halo for the environment creation analysis. This spatial cut is necessary for this analysis because (i) a large fraction would be classified as in the infall region since they did not have time to arrive in the main halo and (ii) we want to understand the assembly process of ICL in protocluster cores. However, there is no reason to not estimate the stellar mass in the infall region at redshift 2 or predict its surface brightness, so we include the infall at redshift 2 in these other analyses. So in the track-back-in-time analysis, we only track back ICL stars classified as being in the main halo at $z=2$, not those in the infall region. However, while these infall region stars are not included in our tracking analysis, we do consider them when estimating the total stellar mass in the infall region, in order to make predictions for future observations.

\vspace{-0.3cm}
\subsection{Galaxy definition}
\label{def_gal}


We use the structure catalogues provided by each simulation to associate the stellar particles with the ICL or with a galaxy. In the Manhattan Suite, the structure catalogue is provided by \textsc{caesar}, while for TNG-Cluster we use the \textsc{subfind} subhalo catalogue. We include only resolved structures with $\rm M_\star \geq 10^{8.5}\,{\rm M_\odot}$, because of the mass resolution of the two simulations mentioned in the previous section and the minimum number of stellar particles in structures for each simulation (10 for Manhattan and 30 for TNG-Cluster).

For TNG-Cluster, we estimate the half-mass radius ($\mathrm{r_{\star, half}}$) of the galaxy considering all the stellar particles inside it. After that, to define the galaxy edges, we multiply this half-mass radius by 2 for satellite galaxies and by 3 for the brightest cluster galaxy. We consider the size of the central galaxy larger than the others because at $z>2$ we can clearly see mass leaking from the central galaxy for $\mathrm{2 \times r_{\star, half}}$, so we decided to have a conservative approach to not overestimate the ICL and multiply it by 3 as in \cite{Kimmig_2025} and \cite{Brown_2026}. For satellite galaxies, we find that the usual literature value of 2 works well for the redshift range considered. This number is supported by previous simulation studies, including TNG, such as \cite{genel_2014, pillepich_2018, Ahvazi_2024, Montenegro-Taborda2025}. The detailed definitions used are described in Table~\ref{table2}.

The Manhattan Suite already provides a list of stellar particles associated with each galaxy previously identified with \textsc{caesar}, in which it considers the spatial distribution of the particles and their dynamics. We decided to use \textsc{caesar} associated particles because it tends to be more robust than a purely spatial association since it considers other dynamical factors. Because the stellar particle selection is already robust, we use the particle \textsc{caesar} lists to associate stellar particles with galaxies or with the ICL. We do not have those lists for TNG-Cluster, so we use the galaxy sizes to select galaxy/ICL stellar particles.

Stellar particles within $\rm R_{200c}$ of the main halo and outside all adopted galaxy apertures are classified as ICL particles. In this first definition, we do not include the ICL particles outside $\mathrm{ R_{200c}}$ at z$\sim$2, because this would bias our estimates of ICL fraction created in the infall region as already mentioned. Including the infall region in the tracking process would result in a high number of particles being classified as created in the infall region, since they were not accreted yet in the main halo. Therefore, we only track particles already in the main halo at $z\sim2$, so we can understand the formation and assembly of ICL that is already in the protocluster core. Although we do not track the ICL particles in the infall region at $z\sim2$, we quantify the total ICL mass inside and outside $\mathrm{R_{200c}}$ as a function of redshift. This is an important prediction that will be useful to observers.

\vspace{-0.5cm}
\subsection{Environment definition}
\label{def_envi}

We define a stellar particle to be in the main halo at a certain redshift if the particle distance from the protocluster center is smaller than its virial radius ($\rm R_{200c}$) at the specific snapshot. A particle is in the infall region if it is outside $\rm R_{200c}$. These definitions can be seen in Fig.~\ref{defs}. We classify ICL particles in all snapshots available in the simulation at $2<z<8$ for Manhattan and at $0<z<8$ for TNG-Cluster.

In particular, we define four different environments, in which one is the main halo and the other three are subcategories of the infall region. More information can be found below.

\begin{figure*}
\centering
\includegraphics[width=2.0\columnwidth]{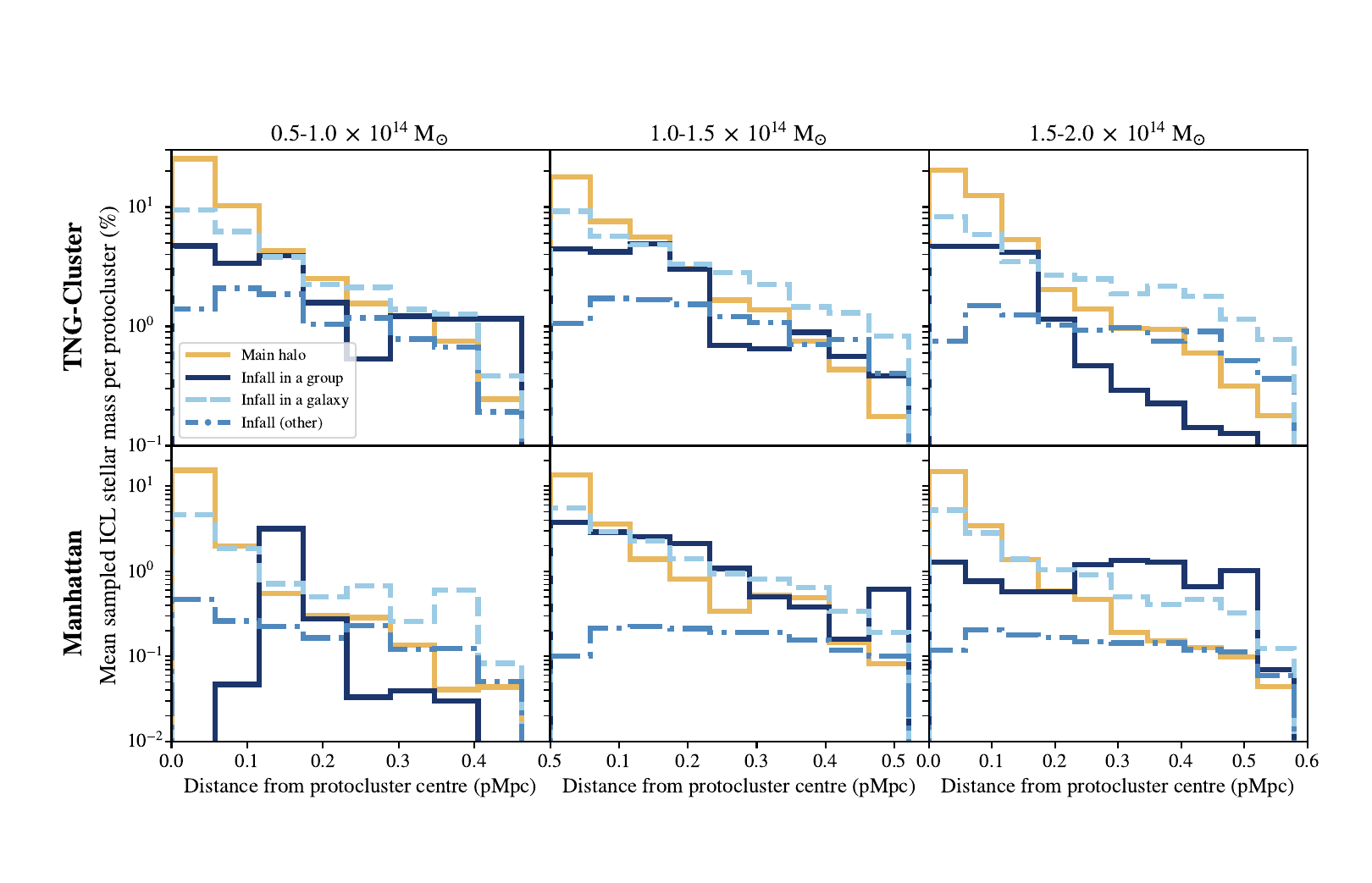}
\vspace{-1.0cm}
\caption{Distances of ICL particles from the cluster centre for different creation environments as defined in Section \ref{def_envi}. The classifications and colours are the same as for Fig.~\ref{piechart}. The y axis shows the mean sampled ICL stellar mass per protocluster in percentages. Top panels show the results for TNG-Cluster, while bottom panels show the results for Manhattan. The three different columns represent the spatial distribution at $z=2$ for protoclusters in different mass ranges. ICL originating in the infall region dominates the outer regions of the main halo, while those originated in the main halo dominate the very inner region. }
\label{dist_bcg}
\end{figure*}

\textbf{Main halo}: The particle is in the main halo if it is inside $\rm R_{200c}$ in the redshift analysed. This classification is represented in orange in Figs.~\ref{piechart}, \ref{infall_def} and \ref{dist_bcg}.

\textbf{Infall in group}: The particle is in the infall in group category if it is outside $\rm R_{200c}$ and it is a member of a halo with $\rm M_{group}\geq10^{13}\,{\rm M_\odot}$ at some point until $z=2$. To define the group edge we define the aperture $\rm R_{\mathrm{group}} \!  = \!  R_{200\mathrm{c},\mathrm{main}}(\mathrm{M}_{\mathrm{group}}/\mathrm{M}_{\mathrm{main}})^{1/3}$ for both simulations. This step is necessary because $\rm R_{200c,group}$ is not available for TNG-Cluster, so we use their masses to estimate their sizes. After that, we select structures inside these groups using these apertures. For Manhattan, the group is selected from \textsc{caesar} halos; for TNG-Cluster, we use the \textsc{subfind} subhalo catalogues. This classification is represented by dark blue.

\textbf{Infall in galaxy}: A particle is classified as infall in a galaxy if it is outside $\rm R_{200c}$ and it is considered as in a galaxy. For TNG-Cluster it has to be inside 2 $\mathrm{r_{\star, half}}$ radius of a galaxy. For Manhattan, the particle has to be listed by \textsc{caesar} as a galaxy member. Structures classified as galaxies must have $\rm M_{200c}\leq10^{13}\,{\rm M_\odot}$, so there is no overlap between galaxies and groups. This classification is represented by light blue.

\textbf{Infall (other)}: The particle is classified as infall (other) if it is in the infall region, but was not assigned to any group or galaxy up to $z=2$. This classification is represented by medium blue.






\vspace{-0.1cm}
\subsection{Galaxy past host definition}
\label{rest}

For each z$\sim$2 ICL particle, we search its past snapshots for associations with resolved galaxies. The last galaxy association before the particle becomes part of the ICL is defined as its last host. We define the first host as the galaxy where the stellar particle was formed. We record both the environment in which the particle was created (creation environment) and the environment in which it left its last identified host (stripping environment). We then estimate the stellar mass for both samples. Stellar masses are provided by \textsc{caesar} for Manhattan and \textsc{subfind} for TNG-Cluster.

\subsection{Stellar particle tracking}
\label{track}

We track back in time our set of ICL particles at $z=2$, and classify them according to their environment. We repeat this process for all snapshots available in the simulation. To find where and when these stars were created, we select the snapshot in which the stellar particle first appeared in the simulation. With that, each particle has a "creation redshift" and a "creation location" associated. Doing that for all ICL particles, we have a distribution of when and where the ICL was first formed according to each simulation. All the results are mass weighted, since particles have different stellar masses, considering that the number of particles alone does not provide the whole picture for ICL assembly.

\vspace{-0.3cm}
\section{Results}

\label{sec:results}

\subsection{ICL formation environment and spatial distribution}
  
As described in Section 3, we track each ICL particle to its formation snapshot and identify the environment, defined in Section \ref{def_envi}, in which it was formed. Our results show that most ICL stars were born in the infall region in both simulations ($\sim$57\% for TNG-Cluster and $\sim$55\% for Manhattan). Of these, most ICL particles were stripped from galaxies, and a small fraction was not associated with any resolved galaxy ($\sim$9\% for TNG-Cluster and $\sim$3\% for Manhattan). These particles, which were formed outside a group or a galaxy, are associated with \textit{in-situ} star formation. Note that there are many definitions of \textit{in-situ} formation in the literature across different astrophysical contexts, so we should take that into account when comparing with literature values. The number of stellar particles that passed through groups in TNG-Cluster (Manhattan) is $\sim18\%$ ($\sim$22\%), which means that several ICL particles were affected by group environments at some point in their lives at high redshifts.

We also analyse how the properties described above vary with the protocluster mass at $z=2$. To do so, we divide the sample into three $ \rm M_{200\mathrm{c}}$ bins: $\rm 0.5\!-\!1.0 \times 10^{14}\,\mathrm{M}_{\odot}$, $\rm 1.0\!-\!1.5 \times 10^{14}\,\mathrm{M}_{\odot}$, and $\rm 1.5\!-\!2.0 \times 10^{14}\,\mathrm{M}_{\odot}$. This allows us to test whether the ICL formation environment depends on protocluster mass.  In Fig.~\ref{piechart}, we show where the ICL particles were born in both simulations, as a function of the mass of the protocluster, and show how these fractions change with the halo mass at $z=2$. Specifically analysing the results of Fig.~\ref{piechart}, we do not find any significant trend related to the protocluster halo mass. Fig.~\ref{piechart} shows that $\rm \sim \! 11.8\!-\!26.8 \%$ of ICL particles have been in groups before joining the main halo at $z=2$, highlighting the relevance of groups in the process of ICL assembly. Groups contribute in mass to the infall region around $28.8-32.8\%$ for TNG-Cluster, while this is $26.0-45.2\%$ for the Manhattan Suite. Note that these values are related to the presence of ICL particles in groups in the past, and not necessarily with ICL stellar stripping within groups.

Fig.~\ref{infall_def} illustrates how the ICL particles are distributed across the protocluster within $R_{200c}$ at $z=2$, in three different projections for one of the TNG-Cluster protoclusters. It is clear that groups have a considerable contribution to the ICL, as shown in dark blue in Fig.~\ref{infall_def}. 

However, it is not possible to see how particles distribute in the very inner regions due to projection effects. With that in mind, we estimate the 3D distances between ICL particles and the cluster center for different types of creation environments, which is shown in Fig.~\ref{dist_bcg}. Particles born in the main halo (orange) end up migrating towards the inner regions of the protocluster, while particles born in the infall region dominate the outer regions ($r\!>\!200$ kpc). Most ICL particles analysed in this work were created in the protocluster infall region ($\rm r>R_{200c}$) and then migrated to their cores later. At the same time, most particles were created inside galaxies and then stripped from them at some point to assemble the ICL. ICL particles from the infall region tend to dominate the outer regions ($r\!>\!200$ kpc) of massive ($\rm 1.5\!-\!2.0 \times 10^{14}\,\mathrm{M}_{\odot}$) protoclusters, compared to protoclusters in the other mass bins. 

As mentioned in Section \ref{sec:TMS}, Fig.~\ref{manhattanproto} shows the spatial distributions of three mass components of protoclusters at $z=2$: stars, dark matter and hot gas in the Manhattan simulation. Their $\rm M_{200c}$ masses are shown in the top right of the first panels. We can clearly see that in most cases, these protoclusters are accreting groups or in the process of merging with another halo of similar size. The hot gas is defined by its temperature ($\mathrm{T\!>\!10^{7} K}$), traced by its emission in the X-rays. We find that the stellar component is clumpier than the hot gas, which is smoother at $z=2$, as shown in Fig.~\ref{manhattanproto}. Some groups surrounding the main halo can be identified through their hot gas emission, making X-ray observations valuable for detecting group-scale structures. This X-ray detection could enable us to observationally study pre-processing in protoclusters and investigate the role of the intragroup light (IGL) in the ICL assembly.

\vspace{-0.3cm}
\subsection{ICL star formation time in protoclusters}

\begin{figure*}
\centering
\includegraphics[width=2\columnwidth]{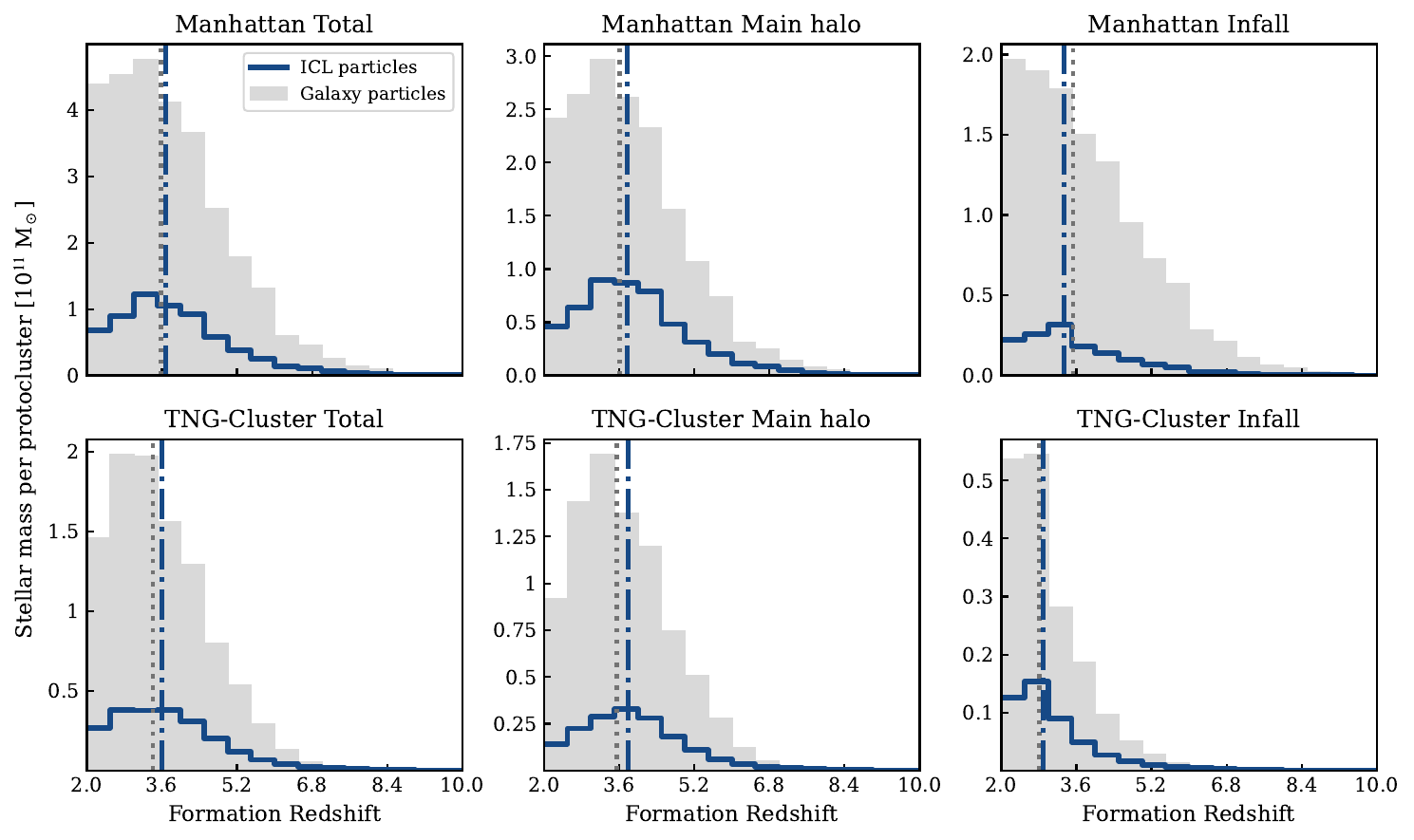}
\vspace{-0.2cm}
\caption{Distribution of formation redshifts for stellar particles in protoclusters. The bottom row shows the results for the TNG-Cluster while the top row is for the Manhattan Suite simulation. For each stellar particle, we associate a redshift in which it first appears in the simulation. The environment is defined as the environment of the ICL particle at $z=2$, and they are classified as in the main halo or infall region. In blue, we show the formation redshift distribution for ICL particles and in grey, we show the formation redshift for stellar particles in protocluster galaxies. The left panel shows the results for the total combined populations considering the infall and main halo environments. The middle panel shows the formation redshift distribution for particles that were already born in the main halo. While the right panel shows the formation redshift distribution for particles born in the infall region. Vertical lines show the median of these distributions, in which the blue dashed lines represent the median for ICL particles, while the dotted gray lines represent the median for galaxy particles. The simulations agree well on the formation redshift of ICL in protoclusters, and both show that main halo ICL particles formed first compared to those in the infall region.}
\label{creation_z}
\end{figure*}

By tracking back all ICL particles to the snapshot in which they formed, we can determine their formation redshift. We present the formation redshift distributions in Fig.~\ref{creation_z}. In this case, we classify particles as in the main halo (middle panels) and in the infall region (right panels) considering their spatial positions at $\rm z\sim2$, while the left panels show the results for the total population combining the main halo and infall region. The colours show the distribution for galaxy stellar particles (grey) and ICL particles (blue). The vertical lines are the median for the two distributions. 

In both simulations, the ICL in the main halo forms earlier than the ICL in the infall regions (Fig.~\ref{creation_z} and Table \ref{tab:main_infall_formation_time_offsets}). The main halo ICL forms at $z_{\rm form}=3.80$ in TNG-Cluster and $3.78$ in Manhattan, compared to $3.89$ and $3.35$ for the infall ICL, respectively. 
Although these average values support the idea that the stellar material contributing to the main halo ICL is systematically older than that associated with infalling structures, the distributions are broad and there are partially overlapping $16$th--$84$th percentile intervals, more information about the mean formation redshift for each environment and the difference in time between them (Gyr) can be found in Table \ref{tab:main_infall_formation_time_offsets}. Specific information about the formation times for stellar particles in galaxies and in the ICL for each environment is presented in Appendix \ref{B} (Table \ref{tab:redshift}).


For the infall region, there is a gap in the median formation time for the two simulations; TNG-Cluster ICL formed later ($\rm z_{form,infall}\sim2.89$) than Manhattan ICL ($\rm z_{form,infall}\sim3.35$). The formation redshifts for the total samples, including both main halo and infall regions (left panels of Fig.~\ref{creation_z}), have similar medians for both simulations. Considering that most of the ICL is concentrated in the main halo at $z=2$, and that we find similar results for the main halo in both simulations, it is expected that the total sample would return similar results in both simulations, with a slight change due to the infall sample.


At a fixed environment, ICL and galaxy particles form at broadly similar epochs. Medians of the distributions are very similar for all cases (total, main halo and infall). In the main halo, median formation redshifts of galaxy and ICL particles differ only modestly, from $3.55$ to $3.80$ in TNG-Cluster and from $3.61$ to $3.78$ in Manhattan. In the infall region, the corresponding values are also comparable, with $\rm z_{\rm form,infall}=2.81$ and $2.89$ in TNG-Cluster and $\rm z_{\rm form,infall}=3.53$ and $3.35$ in Manhattan. The overlap between the 16th--84th percentile ranges further indicates that ICL particles form approximately contemporaneously with galaxy particles in their respective environments.

\begin{table}
\centering
\renewcommand{\arraystretch}{1.6}
\begin{tabular}{l@{\hspace{0.15cm}}c@{\hspace{0.15cm}}c@{\hspace{0.15cm}}c@{\hspace{0.15cm}}c@{\hspace{0.15cm}}c}
\hline
Simulation &
Component &
$\rm z_{\rm form,main}$ &
$\rm z_{\rm form,infall}$ &
$\rm \Delta z_{\rm form}$ &
$\rm \Delta t_{\rm form}$ \\
&
&
&
&
&
$(\mathrm{Gyr})$ \\
\hline
Manhattan & Galaxies & $3.61^{+1.36}_{-1.05}$ & $3.53^{+1.61}_{-1.06}$ & 0.08 & 0.05 \\
Manhattan & ICL & $3.78^{+1.27}_{-1.02}$ & $3.35^{+1.46}_{-0.84}$ & 0.43 & 0.25 \\
TNG-C & Galaxies & $3.55^{+1.23}_{-0.91}$ & $2.81^{+0.98}_{-0.55}$ & 0.74 & 0.54 \\
TNG-C & ICL & $3.80^{+1.16}_{-1.01}$ & $2.89^{+1.04}_{-0.58}$ & 0.91 & 0.60 \\
\hline
\end{tabular}
\vspace{0.2cm}
\caption{
Median formation redshifts of particles formed in the main halo and in the infall region, for galaxy and ICL particles. Quoted ranges in redshift columns correspond to the 16th--84th percentile intervals of the formation-redshift distributions. 
The redshift offset is defined as $\Delta z_{\rm form}=z_{\rm form,main}-z_{\rm form,infall}$. 
The time offset is defined as $\Delta t_{\rm form}=t_{\rm form,infall}-t_{\rm form,main}$. 
}
\label{tab:main_infall_formation_time_offsets}
\end{table}

\vspace{-0.3cm}
\subsection{Progenitor galaxies of the ICL}

\begin{figure*}
\centering
\includegraphics[width=0.6\textwidth, trim= 0 0 150 0,
    clip]{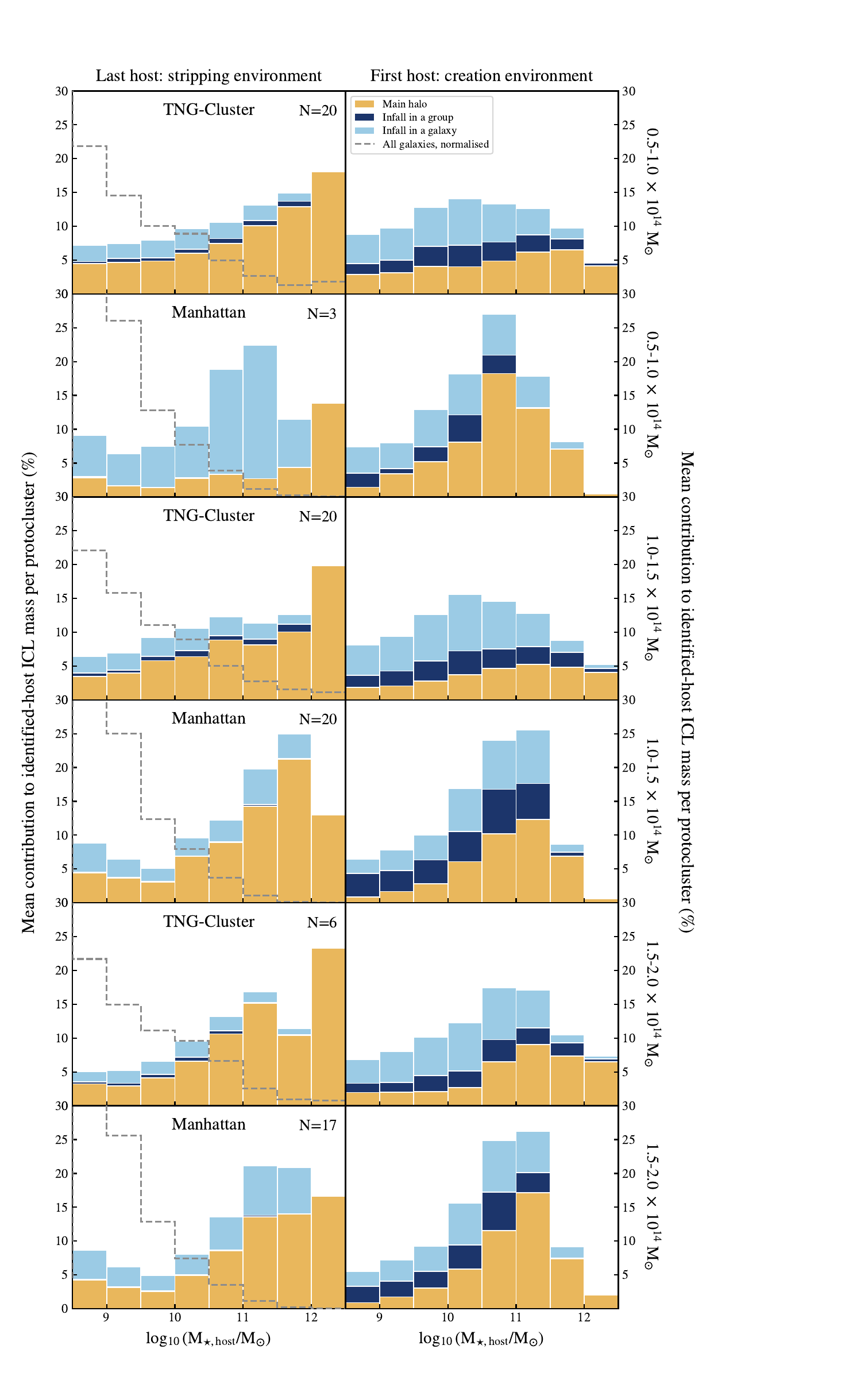}
\vspace{-1.0cm}
\caption{Mass distributions of past host galaxies in different environments for the Manhattan and TNG-Cluster simulations. Each ICL particle identified at $z=2$ that was formed in a galaxy in the past is added to these histograms. The gray dashed histograms show the normalized number of galaxies per protocluster for all protocluster galaxies at $z=2$, the different rows are associated with different mass bins. The left panels show the environments in which the ICL stars were stripped from the galaxies (last host),  while the right panels represent the environment in which the particle was created. Most ICL particles are stripped from galaxies in the main halo, but most of the ICL particles were formed in the infall region. Note that by “stripping” we do not refer exclusively to tidal stripping; rather, we include any physical mechanism capable of removing a stellar particle from a galaxy. We also note that the histograms are stacked rather than overlapping.}
\label{fig:host}
\end{figure*}

We investigate the properties of progenitor galaxies of ICL particles. In particular, we predict the stellar mass distribution of their past hosts. This is particularly relevant because it is a quantity that can be directly compared with observations. For example, one could use ICL observed colors (or stellar populations) to infer its past hosts by selecting galaxies with similar colors and compare their stellar masses with these theoretical predictions. To do that, we use the three samples defined before: (i) main halo; (ii) infall in a galaxy; and (iii) infall in a group. We do not include infall (other) because by definition the particles in this sample did not have a progenitor. In Fig.~\ref{fig:host}, we show the stellar mass distribution of host galaxies in these three environments (orange, light blue, and dark blue, respectively) and also the normalized distribution for all galaxies in the simulation at $z=2$ (gray). Left panels show the distribution of mass of the last host, while right panels show the first host, when the ICL particle formed. We also separate in different mass bins, as represented by the different rows. First hosts tend to be less massive than last hosts in most cases, as reported in Table \ref{tab:first_last_host_mass_comparison_linear}. 

Fig.~\ref{fig:host} clearly illustrates that the mass distribution of ICL progenitors deviates significantly from the protocluster mass function. We can see that most ICL stars were stripped from massive galaxies ($\mathrm{>10^{10.5}\,M_{\odot}}$) already in the main halo, while most of the ICL was created in the infall region and later migrated to the central regions. 
These results are consistent with previous results using the HORIZON-AGN simulation \citep{Brown_2024}, in which they also found that massive galaxies ($\rm >10^{11}\,M_{\odot}$) were the main progenitors of ICL stars. 

We also investigate the dependency of the progenitor host stellar mass with protocluster mass, $\rm M_{200c}$. Differences between samples are very small, and we do not find any strong dependency on protocluster mass. Comparing the two simulations, we find that the average past host stellar masses are similar for both simulations, but TNG-Cluster tends to have a broader distribution compared to Manhattan. Moreover, first and last protocluster mass bins are affected by poor statistics, in which we only have 6 protoclusters for the most massive mass bin for TNG-Cluster and 3 in the last massive bin for Manhattan.

\begin{figure*}
\includegraphics[width=2\columnwidth]{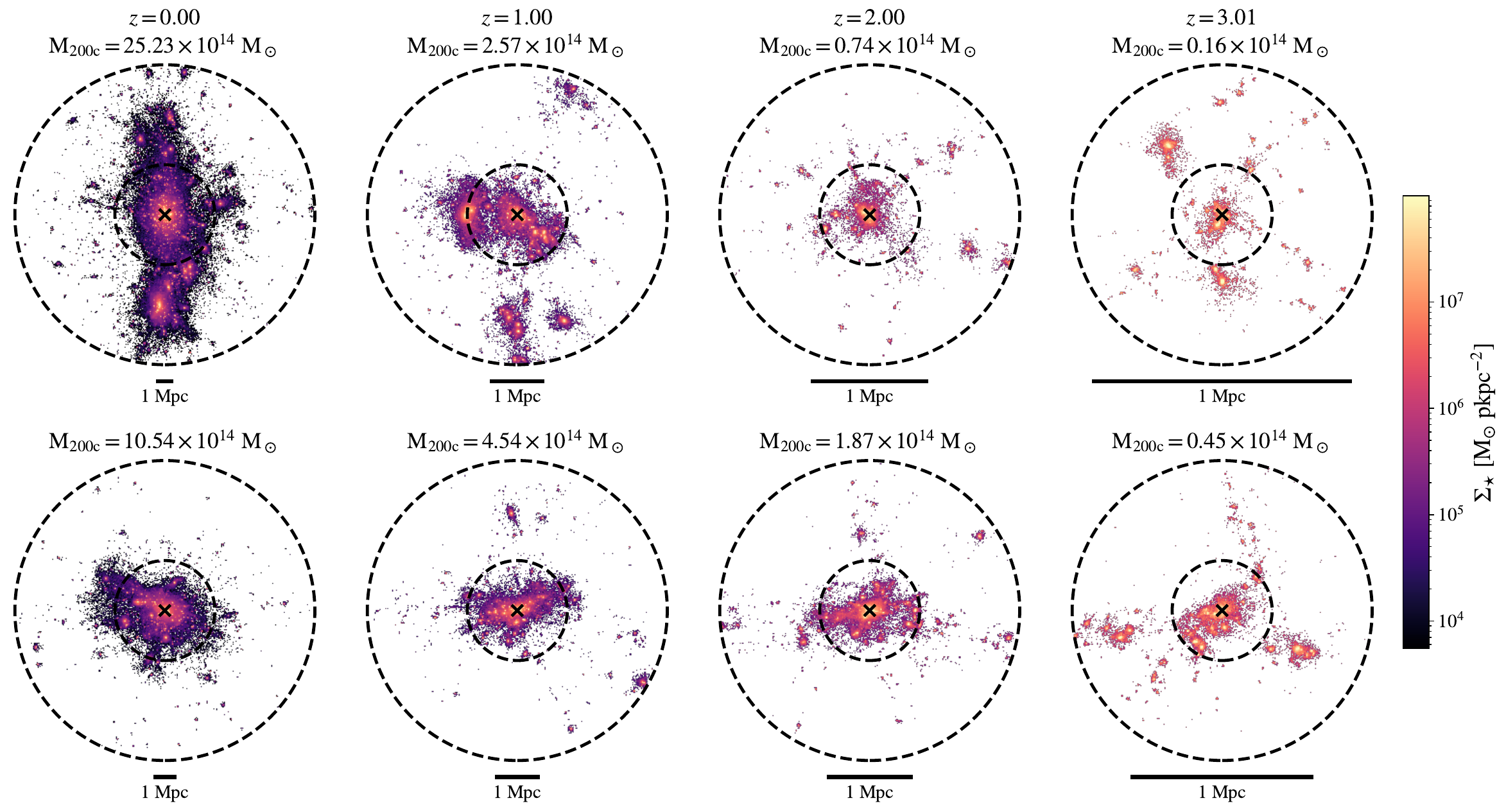}
\vspace{-0.2cm}
\caption{The evolution of the stellar component in the most massive protocluster in TNG-Cluster (bottom panel) and the most massive cluster (upper panel). The inner (outer) circle represents $\rm R_{200c}$ ($3\!\times\! \rm R_{200c}$). The different panels show the stellar distributions at different redshifts, and $\rm M_{200c}$ is also given. The cross marks the center of the protocluster in each panel, 1 Mpc scale bars can be seen in the bottom. The colour bar shows the stellar mass surface density in $\rm M_{\odot} \cdot pkpc^{-2}$. The most massive cluster at redshift 0 is not necessarily the most massive protocluster at $z=2$, as can be seen. Protoclusters are very commonly going through merger events that can move the ICL stars around.}
\label{ICL_tng}
\end{figure*}

\begin{figure}
\includegraphics[width=1\columnwidth]{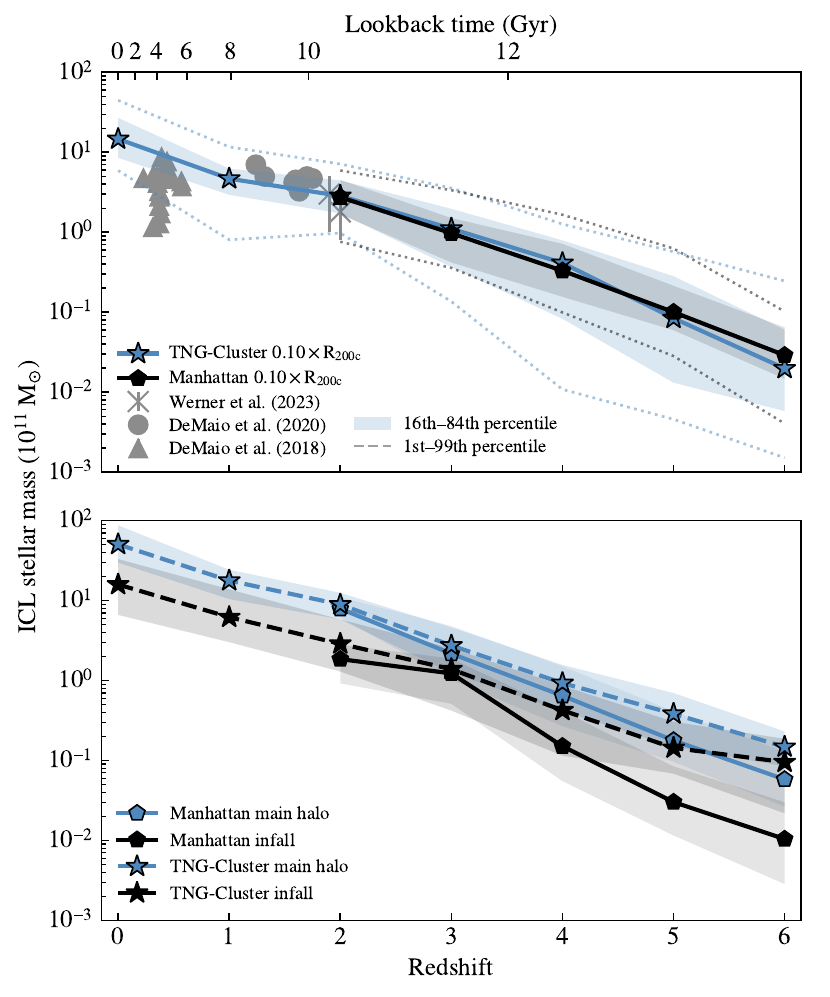}
\vspace{-0.7cm}
\caption[]{ICL stellar mass as a function of redshift for the full selected sample of protoclusters in TNG-Cluster and Manhattan. The upper panel shows the results for an aperture of  $\mathrm{0.10 \! \times \!  R_{200c}}$ and observed values from the literature. The shaded regions show the 16th and 84th percentiles, while the dotted lines show the 1st and 99th percentiles. The bottom panel shows the results for the full main halo and infall regions. The ICL mass was estimated using the method described in Section \ref{rest}, in which particles outside the main halo at z$\sim$2 are also taken into account. In the top panel, the black line represents the results for Manhattan in an aperture of $\mathrm{0.10 \! \times \!  R_{200c}}$ within the main halo and the blue line shows the results for TNG-Cluster. The same panel also shows the results from previous observational studies.  Gray crosses, circles, and triangles represent the observed ICL stellar mass from \cite{DeMaio_2018}, \cite{Demaio2020}, and \cite{Werner_2023}, respectively. The bottom panel shows the total stellar mass of the ICL for the whole main halo and infall region as a function of redshift for both simulations. Stars represent our results for TNG-Cluster (main halo and infall) and pentagons, the results for Manhattan. Note that Manhattan only provides information until $z=2$.}
\label{ICL_obs}
\end{figure}

\begin{table}
\centering
\renewcommand{\arraystretch}{1.6}
\begin{tabular}{l@{\hspace{0.15cm}}c@{\hspace{0.15cm}}c@{\hspace{0.15cm}}c@{\hspace{0.15cm}}c}
\hline
Simulation &
$\rm \left\langle M_{\star,\rm first} \right\rangle$ &
$\rm \left\langle M_{\star,\rm last} \right\rangle$ &
$\rm \Delta M_{\star,\rm host}$ &
$\rm M_{\star,\rm last}/M_{\star,\rm first}$ \\
&
$(10^{10}\,\mathrm{M}_{\odot})$ &
$(10^{10}\,\mathrm{M}_{\odot})$ &
$(10^{10}\,\mathrm{M}_{\odot})$ &
\\
\hline
TNG-C     & 1.34 & 4.07 & 2.74 & 3.05 \\
Manhattan & 1.90 & 8.46 & 6.56 & 4.45 \\
\hline
\end{tabular}
\vspace{0.2cm}
\caption{
Stellar masses of the first and last identified hosts of ICL particles. Last hosts tend to be 3-4 times more massive than last hosts.}
\label{tab:first_last_host_mass_comparison_linear}
\end{table}

\begin{figure}
\includegraphics[width=1\columnwidth]{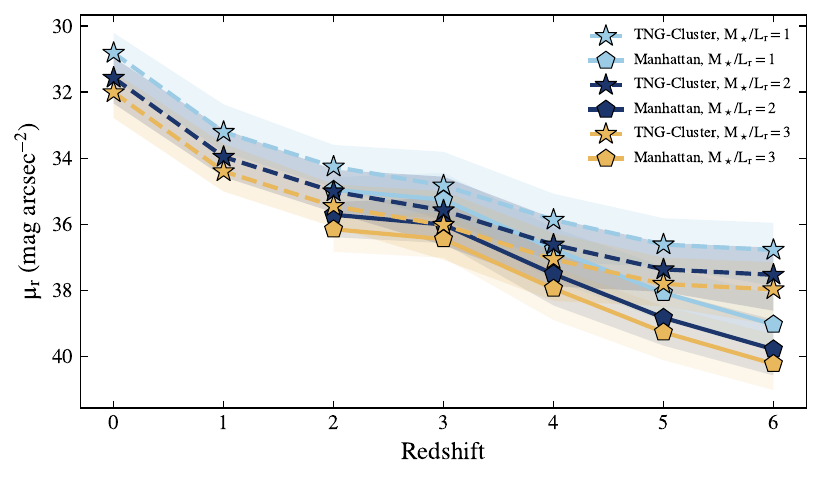}
\vspace{-0.7cm}
\caption[]{Rest-frame r-band surface brightness predictions for infall ICL as a function of redshift for the full sample of protoclusters analysed in this work, including cosmological dimming. The different colours represent different mass-to-light ratios ($\rm M_\star/L=1,2,3$) for different stellar populations, shown in light blue, dark blue, and orange, respectively. The markers represent the two simulations, with TNG-Cluster shown as stars and Manhattan as pentagons. Manhattan is represented by solid lines, while TNG-Cluster is represented by dashed lines. The shaded regions represent the 16th and 84th percentiles.}
\label{brightness}
\end{figure}

\subsection{\boldmath Comparison with observations at 0<z<2}

The goal of this work is to provide theoretical predictions of the ICL in protoclusters that can be easily compared with observations. One of the predictions that we can already test is the amount of mass in the ICL at different redshifts. To do so, we use the full sample of protoclusters already selected and estimate their ICL masses at different redshifts. Fig.~\ref{ICL_tng} shows the distribution of stellar particles in different snapshots for two TNG-Cluster protoclusters as examples. This clearly shows the several groups that compose protoclusters at high redshifts. 

To ensure a fair comparison with observations, we limit our measurements to a maximum outer radius This is necessary because our standard definition includes all ICL within $\rm R_{\mathrm{200c}}$, a radius significantly larger than the limits in observational studies. We chose not to use a fixed maximum radius as it would not be consistent with the definition of the main halo adopted in this paper. At high redshift, such an aperture would not necessarily correspond to the protocluster core region. Considering that $\rm R_{200c}\!\sim\!579$ kpc for TNG and $\rm R_{200c}\!\sim\!572$ kpc for Manhattan in these protoclusters at $z=2$, we decide to explore a maximum radius range around $\sim$30-100 kpc at this redshift. Therefore, we estimate the ICL stellar mass within 0.10$\times \mathrm{R_{200c}}$ of the systems in each snapshot. To remove galaxies, we apply the same galaxy definitions as in Table \ref{table2}, and only select particles not associated with galaxies. To remove the brightest cluster galaxy (BCG) in the TNG simulation, we use an aperture of $\rm 3 r_{\star,half}$, while for Manhattan, we use the membership list generated by \textsc{caesar}, that considers stellar particle position and dynamics. The measured ICL mass evolution with redshift, for both simulations, is shown in Fig.~\ref{ICL_obs}.

We compare our simulation results with estimates of the ICL stellar mass from previous works in clusters at $0.2 \! < \! z \! < \! 2.0$ from \cite{DeMaio_2018, Demaio2020}, and two protoclusters at $z\!\sim\!2$ from \cite{Werner_2023}. We find that most observed measurements match the simulation results in the range of the 1st-99th percentiles, as can be seen in Fig. \ref{ICL_obs}. The observations at $z>1$ match the simulated results within the 16th-84th percentiles. The Manhattan Suite measurements can only be compared to \cite{Werner_2023} results because its lowest redshift snapshot is 2. While we cannot compare with many observations, Manhattan Suite values are comparable to observations at $z=2$ ($\rm M_{\star}\! \sim \! 2.7 \times 10^{11}\,M_{\odot}$ in Manhattan for $\rm 0.10\!\times\! R_{200c}$, $\rm M_{\star} \! \sim \!  1.8-3.0 \times 10^{11}\,M_{\odot}$ in observations). We also highlight that Manhattan values are extremely similar to TNG-Cluster values for the inner regions (top panel of Fig. \ref{ICL_obs}). The CARLA\,J1018 protocluster does not have measured $\rm M_{200c}$ yet, but recently, \cite{Scofield_2026} estimated the mass of XLSSC\,122 finding $\rm M_{200c}\sim1.6\times 10^{14} M_{\odot}$, which is inside the mass range of protoclusters selected for Fig. \ref{ICL_obs} ($\rm M_{200c}\sim 0.5-2.0 \times 10^{14} M_{\odot}$), this could explain the good agreement in ICL mass in the very central region at redshift 2.


TNG-Cluster provides information at $0<z<2$, so we can compare with several observational measurements. 
This simulation match most of the observations we have, although there is a lot of scatter mostly at lower redshifts. This is expected considering observations span a large range of masses and difference in methodologies to extract the ICL mass. For example, different masking techniques (surface brightness cut, galaxy aperture) or the radial extent from the BCG considered.



We see that both simulations suggest that the ICL mass increases steadily until lower redshifts in the central regions (0.10$\times \mathrm{R_{200c}}$), and also in the main halo and infall regions. In the bottom panel of Fig.~\ref{ICL_obs}, we compare the evolution of the ICL stellar mass as a function of redshift, considering all particles within $R_{200c}$ (the main halo) and within $1-3R_{200c}$ (the infall region). As the protocluster grows, the ICL builds up in both environments. 
There is a good agreement between the ICL evolution in the infall region of the two simulations at $z=2$. 
At higher redshifts, TNG-Cluster tends to have more ICL than Manhattan in all environments. At $z<2$, there is a difference in the rate of growth between the main halo and the infall region of the cluster, where the later appears to grow more rapidly. This result echoes the findings in \citet{Demaio2020}, that the ICL at later times grows preferentially at large radiis.

\subsection{Predictions from simulations: ICL at $\mathbf{z>2}$ and in the infall region}


As explained before, we anticipate to find ICL in the infall regions of protoclusters, motivating the observational search for ICL in these regions and at $2<z<8$. As can be seen in Fig.~\ref{ICL_tng}, the ICL in protoclusters is much more extended than measured in observations.  In simulations, the total ICL stellar mass, even when only focusing on the main halo ($<\mathrm{R_{200c}}$, $\sim$ 500 kpc at $z\!\sim\!2$), is much higher than typical measurements in observations, that usually measure the diffuse light until 100-500 kpc from the BCG centre. In this work, we predict that ICL is present in a much more extended volume in clusters ($z<2$) and protoclusters ($z>2$), and can extend out to $\rm 3 \! \times \! \,R_{200c}$ ($\sim$1.5 Mpc at $z=2$). Our results motivate the observational quest to obtain ICL measurements in the outer regions of main haloes. We need deeper data to extend our search to higher redshifts and to regions where the ICL signal-to-noise ratio is not as strong as in the protocluster center (see Fig.\ref{brightness}). 

In Fig.~\ref{brightness}, we show our predictions for the ICL surface brightness in the infall region as a function of redshift. To estimate the expected rest-frame ICL surface brightness in the infall region, we convert the simulated ICL stellar mass into luminosity surface density using three values of stellar mass-to-light ratio ($\rm M_\star/L=1,2,3$), spanning the range of stellar populations that have been observed in the ICL \citep[e.g.][]{DeMaio_2015, Montes_2018}. Note that these M/L values are related to low redshift studies and associated with older stellar populations, but we adopt these also for $z>2$ as a conservative approach. Thus, the ICL can be easier to detect if the ICL has a younger stellar population.

For each redshift, we take the ICL stellar mass in the infall region and distribute it over the projected area of an annulus between $\rm R_{200\mathrm{c}}$ and $\rm 3\!\times\!R_{200\mathrm{c}}$. This gives us a stellar mass surface density, $\Sigma_\star$, in units of $\mathrm{M_\odot\,pc^{-2}}$. We then convert this quantity into luminosity surface density through $\rm \Sigma_L=\Sigma_\star/(M_\star/L)$, where $\rm M_\star/L$ is the assumed stellar mass-to-light ratio in the r band. The luminosity surface density is then converted into rest-frame surface brightness using $\rm \mu = M_{\odot,\mathrm{r}} + 21.572 - 2.5\log_{10}(\Sigma_L)$, where $\rm M_{\odot,\mathrm{r}}$ is the solar absolute magnitude in the r band and the constant 21.572 converts $\mathrm{L_\odot\,pc^{-2}}$ into $\mathrm{mag\,arcsec^{-2}}$. Finally, when estimating the observed surface brightness, we include cosmological dimming by adding $\rm 10\log_{10}(1+z)$ to the rest-frame value. This calculation does not include K-corrections, PSF convolution, background noise, or detailed stellar population modeling, and should therefore be interpreted as an approximate estimate rather than a full mock observation.


The surface brightness of the ICL can vary from 30 $\mathrm{mag\cdot arcsec^{-2}}$ at $\rm z\!\sim\!0$ to 46 $\mathrm{mag\cdot arcsec^{-2}}$ at $\rm z\!\sim\!8$. Surface brightnesses obtained in both simulations are very similar and only differ slightly due to the different M/L ratios assumed. Given the depth of past observations \citep[e.g.][]{Montes_2018, Montes_2022b, Kluge_2025, Canepa2026}, it is possible to currently observe the ICL in the infall region at $z<1$. Note that we are averaging across the entire infall region, between 1 to 3$\rm \times R_{200c}$. If there are any structures or gradients present that concentrate the mass, such as groups and filaments, and therefore increase the surface brightness of some regions, it might be easier to reach higher redshifts. We highlight that the surface brightness values presented in this work are pessimistic estimates, since we are averaging over all the infall region.



%






\section{Discussion}

\label{sec:discussion}

\subsection{\boldmath Impact of galaxy and group definition}

Cross-comparing observational ICL studies remains challenging without a homogeneous measuring framework. Methodological choices, in particular how satellite galaxies are masked and whether and how the BCG is separated from the ICL, can significantly affect the stellar mass budget attributed to the diffuse component. Even though some works consider the BCG+ICL as a single system, this only partially solves the problem, because we still have to excise satellite galaxies (via masking in observations or extracting galaxy particles in simulations). Therefore, in both observational and theoretical works, the definition of a galaxy, i.e., where it ends, makes comparisons between results extremely difficult.

Here, we use different definitions for galaxies in each simulation, as described in Section~\ref{def_gal}, and these definitions can affect the results presented. However, we find a good agreement with observations and between simulations as can be seen in Fig.~\ref{ICL_obs}. This suggests that we have a robust galaxy definition.


In Appendix \ref{A}, we test how our galaxy definition can affect our results for the ICL in the TNG-Cluster simulation. Although quantitative results might change, qualitative results (significant fraction of ICL in the infall region) still hold regardless of the definition. In addition, there are several previous simulation studies that use similar definitions and have consistent results \citep{genel_2014, pillepich_2018, Ahvazi_2024, Montenegro-Taborda2025}.


\subsection{\textit{In-situ} formation and unresolved galaxies}

The relevance of \textit{in-situ} formation in the ICL context is still highly debated. To complicate things even more, there are many definitions of \textit{in-situ} in astronomy. In general, \textit{in-situ} means ``in the natural or original position or place, in the original place instead of being moved to another place''. Conversely, \textit{ex-situ} particles are those particles created in galaxies which then moved to the ICL at some point. While some works suggest \textit{in-situ} formation is negligible \citep{Melnick_2012, DeMaio_2018, Brown_2024}, some other works found it might be up to $\sim$30\% \citep{Ahvazi_2024, Puchwein_2010}. The issue is that different works use different definitions, which makes comparisons extremely hard. In this work, we defined \textit{in-situ} as those particles that were created outside a galaxy and have never been in a galaxy in their life time. We decided to have this conservative approach because the particle could be created in the near region of a galaxy and join the galaxy later, and in these cases it would be considered \textit{in-situ}, although this is just a matter of galaxy edge definition. By specifying that the particle has never been in a galaxy in any snapshots, we ensure that the galaxy definition will not highly affect our \textit{in-situ} fraction.


Interpretation of the results should be taken carefully considering the limitations of the simulations. Currently, there are few observational measurements of \textit{in-situ} star formation \citep{Melnick_2012, Barfety_2022}. However, \cite{Melnick_2012} report <1\% for \textit{in-situ} star formation, while \cite{Barfety_2022} find between 15-21\%. Our measurements suggest that this fraction varies between 3-10\%, as can be seen in medium blue colour of Fig.~\ref{piechart}. Note that in this work an ICL particle is classified as formed “in-situ” or “infall (other)” if they are (i) classified as ICL at $z=2$, (ii) was born in the infall region, and (iii) has never been in a galaxy or a group in their whole life.

However, it includes several possible scenarios: (i) star formation in gas stripped from galaxy, (ii) stars that are outside the galaxy definition, (iii) stars in a strongly disrupted system and (iv) a low mass galaxy that is not identified by the simulation. We also note that the \textsc{caesar} galaxy definition used in Manhattan seems to be more robust compared to the spatial definition we used for TNG-Cluster. We see that in the infall (other) classification (Fig.~\ref{piechart}),  more ICL particles were assigned to galaxies in Manhattan (97\%) than in TNG-Cluster (91\%).


\section{Conclusions}
We used the Manhattan Suite and TNG-Cluster simulations to investigate the formation and assembly of ICL in protoclusters. We defined ICL particles at z$\sim$2 and tracked them back in time. A summary of our main findings is described below:



\begin{itemize}
    \item[--] There is a considerable amount of ICL in the infall region of protoclusters at 2<z<4, ranging from $\rm 0.1\times10^{11} \! < \! M_{*}/M_{\odot} \! < \! 3\times10^{11}$. At redshift 0, this mass can reach $\rm \sim \! 1.6\times 10^{12}M_{\odot}$.
    \item[--] ICL stars in protoclusters at $z=2$ formed mainly in the infall region, in which $\sim$55\% of stellar particles were created in this region for Manhattan and $\sim\!$57\% for TNG-Cluster. Of these, a considerable amount of ICL stars were associated with groups, with mean values of $\sim$12-28\% depending on the simulation, environment and mass of the protocluster. 
    \item[--] Most ICL stars were stripped from massive galaxies ($\sim10^{10.5}\mathrm{M}_{\odot}$) in the main halo, while most of these stars were born in the infall regions of protoclusters ($\sim56\%$). Galaxies where ICL stars were born are 3--4 times less massive than galaxies where they were stripped, which supports the ideas that: (i) galaxies grew in mass with time and/or (ii) stars migrated to a more massive galaxy before being stripped. The mass function of galaxies that hosted ICL stars in the past tends to be composed of more massive galaxies compared to the overall population of protocluster galaxies at $z=2$. This effect is even stronger if galaxies were initially in groups. 
    \item[--] ICL stars in different environments were born in different cosmic epochs. The formation redshift distribution for each environment follows the general stellar population, meaning that ICL stars formed at the same time as stars inside galaxies if they are in the same environment. This means that an ICL particle in the main halo was formed at the same time as a particle in a galaxy that is also in the main halo. However, these formation times change for different global environments, in which main halo particles were formed prior to the infall particles.
    \item[--] We predict that the rest-frame surface brightness in the r band for the ICL in the infall region can vary from 30 $\mathrm{mag\,arcsec^{-2}}$ at $\rm z\!\sim\!0$ to 46 $\mathrm{mag\,arcsec^{-2}}$ at $\rm z\!\sim\!8$, depending on the mass-to-light ratio and redshift considered. These values are approximate pessimistic estimates based only on the projected ICL stellar mass over the whole infall region, assumed mass-to-light ratios, and cosmological surface brightness dimming. 
    \item[--] Stars formed in the main halo occupy the central regions of the protocluster at z$\sim$2, while stars formed in the infall dominate in the outer regions, as expected by the hierarchical assembly of structures.  
    
\end{itemize}

More observations are necessary to constrain these simulated results. In particular, we should (i) investigate the infall region of protoclusters and look for ICL, in particular IGL in groups; (ii) measure the ages of ICL stellar populations in the infall and main halo to constrain the formation times predicted by the simulations; (iii) explore possible ICL host galaxies in the early Universe, and (iv) look for ICL at high redshifts. Deep observations with JWST, together with next-generation telescopes such as Euclid \citep{Euclid_2025} and Roman \citep{Spergel_2013}, and dedicated processing techniques, will provide deeper imaging of protoclusters and enable more direct constraints on simulations.


\label{sec:conclusions}


\vspace{-0.5cm}
\section*{Data Availability}
The TNG-Cluster simulation used in the analysis of this paper can be downloaded from the TNG-Project website (\href{https://www.tng-project.org/cluster/}{https://www.tng-project.org/cluster/}). The Manhattan Suite simulation can be accessed through the Flatiron Institute's facilities by contacting the authors.

\vspace{-0.5cm}

\section*{Acknowledgements}

This work was supported by collaborative visits funded by the Cosmology and Astroparticle Student and Postdoc Exchange Network (CASPEN). SW and MJ acknowledge support from the United Kingdom Research and Innovation (UKRI) Future Leaders Fellowship `Using Cosmic Beasts to uncover the Nature of Dark Matter' (grant number MR/X006069/1). SW acknowledges Dr. Annalisa Pillepich, Dr. David Lagattuta, Dr. Jess Doppel, Dr. Benjamin Beauchesne, and Dr. Ivan Almeida for helpful discussions during this work.

The simulations presented in this work were run on the Flatiron Institute's research computing facilities (the Iron compute cluster), supported by the Simons Foundation.  DR thanks Romeel Dav\'{e} for providing the \textsc{Simba} code for this project and for helpful discussions, and Ulrich P. Steinwandel for the parent N-body simulation.  DR additionally thanks Connor Bottrell, and Asya Rennehan for insightful discussions during the course of this research.

Y.J-T. acknowledges financial support from the State Agency for Research of the Spanish MCIU through Center of Excellence Severo Ochoa award to the Instituto de Astrofísica de Andaluc\'ia CEX2021-001131-S funded by MCIN/AEI/10.13039/501100011033, and from the grant PID2022-136598NB-C32 Estallidos and project ref. AST22-00001-Subp-15 funded by the EU-NextGenerationEU.

MM acknowledges support from grant RYC2022-036949-I financed by the MICIU/AEI/10.13039/501100011033 and by ESF+, grant PID2024-158845NB-I00  financed by MICIU/AEI/10.13039/501100011033 and by ERDF, EU, and program Unidad de Excelencia Mar\'{i}a de Maeztu CEX2020-001058-M.

\textit{Facility}: Rusty (Flatiron Institute), TNG-Cluster API (Max Planck Computing \& Data Facility; FAS Research Computing Harvard University)

\textit{Software:} Our analysis was performed using the Python programming language. The following packages were used throughout the analysis: \textsc{h5py} \citep{Collette2013}, \textsc{numpy} \citep{Harris2020}, \textsc{scipy} \citep{Virtanen2020}, \textsc{yt} \citep{Turk2011}, and \textsc{matplotlib} \citep{Hunter2007}. Prototyping of the analysis scripts was performed in the IPython environment \citep{Perez2007}.

We thank the anonymous referee for their detailed and constructive comments, which significantly improved the quality of this manuscript.
\vspace{-0.5cm}



\bibliographystyle{mnras}
\bibliography{bibliography.bib} 

\clearpage 
\onecolumn
\appendix

\section{\centering Statistics for ICL star formation epochs, environments and past hosts}
\label{A}

We attach specific tables for the results in this paper. In Tables \ref{tab:tng_cluster_sample_summary} and \ref{tab:manhattan_sample_summary}, we give a summary of the protoclusters used in this work. In Table \ref{tab:icl_piechart_pathway_fractions} we give details about the environment classifications for different mass bins in each simulation. In Table \ref{tab:redshift} we give information about the formation epochs for ICL and galaxy stellar particles. Finally, in Table \ref{tab:hosts} we give information about the past hosts analysis.

\begin{table}[H]
\centering
\renewcommand{\arraystretch}{1.5}
\scriptsize
\setlength{\tabcolsep}{4.0pt}
\begin{tabular}{cccccccc}
\hline
Mass bin & Cluster ID & $\rm M_{200\mathrm{c}}$ & $\rm R_{200\mathrm{c}}$ & Main & Infall (other). & Infall in gal. & Infall ever group  \\
 &  & $[10^{14}\,\mathrm{M}_{\odot}]$ & $[\mathrm{pMpc}]$ & $[\%]$ & $[\%]$ & $[\%]$ & $[\%]$ \\
\hline
1 & 140 & 0.716 & 0.420 & 44.57 & 6.55 & 18.38 & 30.49 \\
1 & 204 & 0.722 & 0.422 & 52.76 & 7.73 & 39.51 & 0.00 \\
1 & 19 & 0.723 & 0.422 & 58.97 & 4.19 & 21.99 & 14.85 \\
1 & 91 & 0.727 & 0.422 & 62.86 & 6.96 & 30.18 & 0.00 \\
1 & 196 & 0.731 & 0.423 & 61.23 & 8.27 & 30.50 & 0.00 \\
1 & 304 & 0.736 & 0.424 & 57.38 & 8.49 & 34.14 & 0.00 \\
1 & 114 & 0.736 & 0.424 & 29.99 & 12.64 & 20.88 & 36.50 \\
1 & 9 & 0.739 & 0.425 & 35.42 & 12.69 & 26.31 & 25.58 \\
1 & 1 & 0.744 & 0.426 & 43.32 & 17.30 & 39.38 & 0.00 \\
1 & 35 & 0.745 & 0.426 & 42.78 & 11.58 & 30.02 & 15.61 \\
1 & 201 & 0.747 & 0.426 & 44.46 & 16.19 & 39.35 & 0.00 \\
1 & 184 & 0.754 & 0.428 & 34.79 & 6.78 & 15.82 & 42.61 \\
1 & 82 & 0.762 & 0.429 & 24.97 & 1.47 & 3.30 & 70.26 \\
1 & 267 & 0.765 & 0.430 & 36.85 & 6.86 & 15.73 & 40.56 \\
1 & 221 & 0.767 & 0.430 & 65.62 & 8.42 & 25.96 & 0.00 \\
1 & 81 & 0.773 & 0.431 & 61.08 & 8.55 & 30.37 & 0.00 \\
1 & 85 & 0.775 & 0.432 & 37.39 & 12.15 & 26.96 & 23.51 \\
1 & 276 & 0.777 & 0.432 & 37.88 & 8.68 & 24.54 & 28.90 \\
1 & 133 & 0.779 & 0.432 & 45.35 & 8.33 & 22.48 & 23.84 \\
1 & 126 & 0.782 & 0.433 & 46.49 & 10.34 & 43.18 & 0.00 \\
\hline
2 & 94 & 1.050 & 0.478 & 52.70 & 7.54 & 32.05 & 7.71 \\
2 & 101 & 1.067 & 0.480 & 32.76 & 9.49 & 25.40 & 32.34 \\
2 & 4 & 1.130 & 0.489 & 56.74 & 10.28 & 32.98 & 0.00 \\
2 & 29 & 1.149 & 0.492 & 35.46 & 3.69 & 7.95 & 52.91 \\
2 & 167 & 1.162 & 0.494 & 25.52 & 9.86 & 33.03 & 31.58 \\
2 & 8 & 1.164 & 0.494 & 18.23 & 10.06 & 22.99 & 48.71 \\
2 & 50 & 1.167 & 0.495 & 44.21 & 3.98 & 16.15 & 35.65 \\
2 & 31 & 1.205 & 0.500 & 32.77 & 17.28 & 44.88 & 5.06 \\
2 & 97 & 1.228 & 0.503 & 40.43 & 7.49 & 31.04 & 21.04 \\
2 & 87 & 1.231 & 0.504 & 28.47 & 3.23 & 8.83 & 59.46 \\
2 & 179 & 1.256 & 0.507 & 43.93 & 5.68 & 17.16 & 33.22 \\
2 & 17 & 1.258 & 0.507 & 47.63 & 12.19 & 40.18 & 0.00 \\
2 & 131 & 1.281 & 0.510 & 20.29 & 14.30 & 49.70 & 15.70 \\
2 & 46 & 1.292 & 0.512 & 47.13 & 12.96 & 39.91 & 0.00 \\
2 & 141 & 1.294 & 0.512 & 51.06 & 11.09 & 37.85 & 0.00 \\
2 & 103 & 1.304 & 0.513 & 36.02 & 12.03 & 29.28 & 22.67 \\
2 & 7 & 1.311 & 0.514 & 35.43 & 9.14 & 33.67 & 21.76 \\
2 & 36 & 1.319 & 0.515 & 39.99 & 10.23 & 42.80 & 6.98 \\
2 & 12 & 1.319 & 0.515 & 42.21 & 15.40 & 42.39 & 0.00 \\
2 & 18 & 1.375 & 0.523 & 38.80 & 16.54 & 44.66 & 0.00 \\
\hline
3 & 43 & 1.514 & 0.540 & 43.37 & 11.61 & 43.89 & 1.13 \\
3 & 158 & 1.527 & 0.541 & 34.22 & 11.90 & 33.58 & 20.31 \\
3 & 37 & 1.650 & 0.555 & 51.07 & 8.10 & 26.36 & 14.47 \\
3 & 0 & 1.684 & 0.559 & 40.60 & 7.18 & 22.03 & 30.19 \\
3 & 14 & 1.723 & 0.563 & 51.54 & 5.76 & 19.64 & 23.06 \\
3 & 106 & 1.868 & 0.579 & 46.01 & 9.12 & 37.85 & 7.02 \\
\hline
\end{tabular}
\vspace{0.1cm}
\caption{
TNG-Cluster sample used in the ICL star formation environment analysis.
For each protocluster, we list the halo mass at $z=2$, $\rm R_{200\mathrm{c}}$, and selected mass weighted fractions of the sampled ICL component.
}
\label{tab:tng_cluster_sample_summary}
\end{table}

\begin{table*}
\centering
\renewcommand{\arraystretch}{1.7}
\scriptsize
\setlength{\tabcolsep}{4.0pt}
\begin{tabular}{cccccccc}
\hline
Mass bin & Cluster ID & $\rm M_{200\mathrm{c}}$ & $\rm R_{200\mathrm{c}}$ & Main & Infall (other). & Infall in gal. & Infall ever group \\
 &  & $[10^{14}\,\mathrm{M}_{\odot}]$ & $[\mathrm{pMpc}]$ & $[\%]$ & $[\%]$ & $[\%]$ & $[\%]$ \\
\hline
1 & 84 & 0.605 & 0.394 & 53.92 & 2.88 & 18.40 & 24.80 \\
1 & 87 & 0.872 & 0.445 & 49.84 & 10.76 & 28.95 & 10.45 \\
1 & 69 & 0.881 & 0.446 & 60.42 & 4.66 & 34.90 & 0.02 \\
\hline
2 & 43 & 1.314 & 0.510 & 20.50 & 2.83 & 37.78 & 38.90 \\
2 & 49 & 1.318 & 0.510 & 43.41 & 4.75 & 29.80 & 22.04 \\
2 & 46 & 1.320 & 0.511 & 55.68 & 4.53 & 39.77 & 0.02 \\
2 & 55 & 1.325 & 0.512 & 39.88 & 3.13 & 26.51 & 30.48 \\
2 & 45 & 1.328 & 0.512 & 60.14 & 2.15 & 37.71 & 0.00 \\
2 & 26 & 1.342 & 0.514 & 13.26 & 2.54 & 16.18 & 68.02 \\
2 & 44 & 1.348 & 0.514 & 36.01 & 4.62 & 40.41 & 18.96 \\
2 & 31 & 1.357 & 0.515 & 22.40 & 2.02 & 25.27 & 50.32 \\
2 & 39 & 1.393 & 0.520 & 47.85 & 2.48 & 24.32 & 25.34 \\
2 & 34 & 1.404 & 0.521 & 35.04 & 3.36 & 13.54 & 48.06 \\
2 & 37 & 1.404 & 0.521 & 35.27 & 2.30 & 33.74 & 28.69 \\
2 & 38 & 1.406 & 0.522 & 53.16 & 2.31 & 39.47 & 5.06 \\
2 & 40 & 1.407 & 0.522 & 35.99 & 3.65 & 34.00 & 26.36 \\
2 & 30 & 1.422 & 0.524 & 47.58 & 2.72 & 23.38 & 26.32 \\
2 & 35 & 1.434 & 0.525 & 49.61 & 3.03 & 39.75 & 7.61 \\
2 & 29 & 1.448 & 0.527 & 64.16 & 2.92 & 32.92 & 0.00 \\
2 & 24 & 1.463 & 0.529 & 63.00 & 3.74 & 33.25 & 0.00 \\
2 & 28 & 1.464 & 0.529 & 17.72 & 2.17 & 15.58 & 64.53 \\
2 & 25 & 1.494 & 0.532 & 42.09 & 3.18 & 29.98 & 24.75 \\
2 & 27 & 1.496 & 0.533 & 32.25 & 1.99 & 16.11 & 49.66 \\
\hline
3 & 19 & 1.516 & 0.535 & 15.47 & 3.34 & 19.13 & 62.07 \\
3 & 20 & 1.538 & 0.537 & 41.52 & 2.98 & 22.61 & 32.89 \\
3 & 22 & 1.550 & 0.539 & 47.68 & 2.99 & 26.02 & 23.31 \\
3 & 23 & 1.558 & 0.540 & 46.82 & 3.31 & 31.50 & 18.36 \\
3 & 21 & 1.604 & 0.545 & 61.51 & 3.37 & 23.35 & 11.78 \\
3 & 18 & 1.615 & 0.546 & 36.07 & 3.50 & 29.27 & 31.16 \\
3 & 15 & 1.628 & 0.548 & 60.52 & 3.17 & 36.31 & 0.00 \\
3 & 17 & 1.654 & 0.551 & 59.93 & 4.11 & 35.96 & 0.00 \\
3 & 16 & 1.662 & 0.552 & 56.20 & 3.89 & 34.22 & 5.69 \\
3 & 14 & 1.688 & 0.554 & 44.78 & 2.58 & 22.55 & 30.10 \\
3 & 11 & 1.852 & 0.572 & 24.39 & 1.77 & 25.99 & 47.85 \\
3 & 9 & 1.906 & 0.577 & 63.57 & 3.06 & 33.37 & 0.00 \\
3 & 12 & 1.918 & 0.579 & 48.36 & 2.78 & 42.59 & 6.26 \\
3 & 10 & 1.919 & 0.579 & 77.27 & 3.39 & 19.34 & 0.00 \\
3 & 33 & 1.935 & 0.580 & 56.06 & 5.18 & 18.87 & 19.89 \\
3 & 8 & 1.937 & 0.580 & 38.61 & 2.24 & 28.68 & 30.48 \\
3 & 13 & 1.938 & 0.581 & 50.34 & 3.64 & 38.67 & 7.35 \\
\hline
\end{tabular}
\vspace{0.5cm}
\caption{
Manhattan Suite sample used in the ICL star formation environment analysis.
For each protocluster, we list the halo mass at $z=2$, $\rm R_{200\mathrm{c}}$, and selected mass weighted fractions of the sampled ICL component. }
\label{tab:manhattan_sample_summary}
\end{table*}


\begin{table*}
    \centering
    \small
    \renewcommand{\arraystretch}{1.3}
    \resizebox{\textwidth}{!}{%
    \begin{tabular}{lllcccc}
        \toprule
        Simulation &
        Protocluster mass range &
        Formation pathway &
        $\rm N_{\rm cl}$ &
        Mean ICL fraction (\%) &
        Median ICL fraction (\%) &
        $\rm \sigma_{\rm protocluster}$ (\%) \\
        \midrule

        \multirow{4}{*}{TNG-Cluster}
        & \multirow{4}{*}{$0.5$--$1.0 \times 10^{14}\ {\rm M}_{\odot}$}
        & Main halo
        & 20
        & $46.2^{+2.5}_{-2.8}$
        & $44.5^{+16.5}_{-9.0}$
        & $11.8$ \\

        &
        & Infall in a group
        & 20
        & $17.6^{+4.4}_{-4.3}$
        & $15.2^{+21.0}_{-15.2}$
        & $19.9$ \\

        &
        & Infall in a galaxy
        & 20
        & $26.9^{+2.1}_{-2.2}$
        & $26.6^{+12.5}_{-8.2}$
        & $9.7$ \\

        &
        & Infall (other)
        & 20
        & $9.2^{+0.8}_{-0.8}$
        & $8.5^{+4.2}_{-1.7}$
        & $3.8$ \\

        \addlinespace[3pt]

        \multirow{4}{*}{TNG-Cluster}
        & \multirow{4}{*}{$1.0$--$1.5 \times 10^{14}\ {\rm M}_{\odot}$}
        & Main halo
        & 20
        & $38.5^{+2.3}_{-2.3}$
        & $39.4^{+8.2}_{-10.7}$
        & $10.3$ \\

        &
        & Infall in a group
        & 20
        & $19.7^{+4.2}_{-4.3}$
        & $18.4^{+17.2}_{-18.4}$
        & $19.4$ \\

        &
        & Infall in a galaxy
        & 20
        & $31.6^{+2.6}_{-2.5}$
        & $33.0^{+9.8}_{-15.6}$
        & $12.0$ \\

        &
        & Infall (other)
        & 20
        & $10.1^{+0.9}_{-0.9}$
        & $10.1^{+4.1}_{-4.4}$
        & $4.1$ \\

        \addlinespace[3pt]

        \multirow{4}{*}{TNG-Cluster}
        & \multirow{4}{*}{$1.5$--$2.0 \times 10^{14}\ {\rm M}_{\odot}$}
        & Main halo
        & 6
        & $44.5^{+2.4}_{-2.6}$
        & $44.7^{+6.5}_{-5.4}$
        & $6.6$ \\

        &
        & Infall in a group
        & 6
        & $16.0^{+4.3}_{-4.1}$
        & $17.4^{+7.1}_{-11.5}$
        & $10.7$ \\

        &
        & Infall in a galaxy
        & 6
        & $30.6^{+3.8}_{-3.5}$
        & $30.0^{+9.1}_{-8.4}$
        & $9.5$ \\

        &
        & Infall (other)
        & 6
        & $8.9^{+1.0}_{-0.9}$
        & $8.6^{+3.1}_{-1.7}$
        & $2.4$ \\

        \addlinespace[3pt]

        \multirow{4}{*}{Manhattan}
        & \multirow{4}{*}{$0.5$--$1.0 \times 10^{14}\ {\rm M}_{\odot}$}
        & Main halo
        & 3
        & $54.7^{+2.2}_{-2.2}$
        & $53.9^{+4.4}_{-2.8}$
        & $5.3$ \\

        &
        & Infall in a group
        & 3
        & $11.8^{+4.8}_{-4.8}$
        & $10.5^{+9.8}_{-7.1}$
        & $12.4$ \\

        &
        & Infall in a galaxy
        & 3
        & $27.4^{+5.5}_{-3.5}$
        & $28.9^{+4.0}_{-7.2}$
        & $8.4$ \\

        &
        & Infall (other)
        & 3
        & $6.1^{+2.0}_{-2.0}$
        & $4.7^{+4.1}_{-1.2}$
        & $4.1$ \\

        \addlinespace[3pt]

        \multirow{4}{*}{Manhattan}
        & \multirow{4}{*}{$1.0$--$1.5 \times 10^{14}\ {\rm M}_{\odot}$}
        & Main halo
        & 20
        & $40.7^{+3.3}_{-3.2}$
        & $41.0^{+14.6}_{-18.2}$
        & $14.9$ \\

        &
        & Infall in a group
        & 20
        & $26.8^{+4.6}_{-4.7}$
        & $25.8^{+23.8}_{-25.6}$
        & $21.3$ \\

        &
        & Infall in a galaxy
        & 20
        & $29.5^{+1.9}_{-2.0}$
        & $31.4^{+8.0}_{-15.0}$
        & $9.0$ \\

        &
        & Infall (other)
        & 20
        & $3.0^{+0.2}_{-0.2}$
        & $2.9^{+0.9}_{-0.7}$
        & $0.9$ \\

        \addlinespace[3pt]

        \multirow{4}{*}{Manhattan}
        & \multirow{4}{*}{$1.5$--$2.0 \times 10^{14}\ {\rm M}_{\odot}$}
        & Main halo
        & 17
        & $48.8^{+3.3}_{-3.5}$
        & $48.4^{+12.6}_{-10.9}$
        & $15.0$ \\

        &
        & Infall in a group
        & 17
        & $19.2^{+4.0}_{-4.3}$
        & $18.4^{+13.6}_{-18.4}$
        & $18.2$ \\

        &
        & Infall in a galaxy
        & 17
        & $28.7^{+1.7}_{-1.7}$
        & $28.7^{+7.4}_{-7.5}$
        & $7.3$ \\

        &
        & Infall (other)
        & 17
        & $3.3^{+0.2}_{-0.2}$
        & $3.3^{+0.4}_{-0.6}$
        & $0.8$ \\

        \bottomrule
    \end{tabular}%
    }
    \vspace{0.5cm}
        \caption{
    Fractional contribution to the ICL stellar mass from each formation
    pathway, divided by protocluster mass at $\rm z\simeq2$.
    The quoted uncertainties on the mean ICL fractions correspond to the
    $16$th--$84$th percentile interval obtained by bootstrap resampling of
    all protoclusters within each mass bin.
    The median interval and $\rm \sigma_{\rm protocluster}$ quantify the
    protocluster-to-protocluster spread.
    }
    \label{tab:icl_piechart_pathway_fractions}
\end{table*}

\begin{table*}
\centering
 \renewcommand{\arraystretch}{1.3}
\begin{tabular}{lllccc}
\toprule
Simulation & Region & Component & $\langle z_{\rm form}\rangle$  & $z_{\rm form,50}$ (16th, 84th) & $\rm \sigma_{z_{\rm form}}$ \\
\midrule
Manhattan Suite & Total & Galaxy particles & $3.79$ &  $3.58^{+1.45}_{-1.06}$ & $1.259$ \\
 & Total & ICL particles & $3.88$  & $3.68^{+1.31}_{-1.00}$ & $1.213$ \\
 & Main halo & Galaxy particles & $3.80$ & $3.61^{+1.36}_{-1.05}$ & $1.222$ \\
 & Main halo & ICL particles & $3.94$  & $3.78^{+1.27}_{-1.02}$ & $1.207$ \\
 & Infall & Galaxy particles & $3.79$ &  $3.53^{+1.61}_{-1.06}$ & $1.314$ \\
 & Infall & ICL particles & $3.63$ &  $3.35^{+1.46}_{-0.84}$ & $1.202$ \\
TNG-Cluster & Total & Galaxy particles & $3.41$ & $3.41^{+1.24}_{-0.87}$ & $1.048$ \\
 & Total & ICL particles & $3.76$  & $3.60^{+1.24}_{-0.99}$ & $1.164$ \\
 & Main halo & Galaxy particles & $3.71$ & $3.55^{+1.23}_{-0.91}$ & $1.048$ \\
 & Main halo & ICL particles & $3.93$  & $3.80^{+1.16}_{-1.01}$ & $1.152$ \\
 & Infall & Galaxy particles & $3.02$  & $2.81^{+0.98}_{-0.55}$ & $0.848$ \\
 & Infall & ICL particles & $3.15$ & $2.89^{+1.04}_{-0.58}$ & $0.989$ \\
\bottomrule
\end{tabular}
\vspace{0.5cm}
\caption{Stellar particles formation redshifts for TNG-Cluster and the Manhattan Suite, computed from the same formation redshift histogram inputs used in Fig.~\ref{creation_z}. The 16th-84th percentile interval quantify the width measurement of each formation redshift distribution.}
\label{tab:redshift}
\end{table*}

\begin{table*}
    \centering
    \small
    \renewcommand{\arraystretch}{1.55}
    \resizebox{\textwidth}{!}{%
    \begin{tabular}{llllcccc}
        \toprule
        Simulation &
        Host analysis &
        Protocluster mass range &
        Host channel &
        $\mathrm{N}_{\mathrm{cl}}$ &
        $\rm \left\langle \log_{10} M_{\star,\mathrm{host}} \right\rangle$ &
        $\rm \mathrm{median}\left(\log_{10} M_{\star,\mathrm{host}}\right)$ &
        $\rm \sigma_{\mathrm{protocluster}}$ \\
        &
        &
        &
        &
        &
        $[\mathrm{M}_{\odot}]$ &
        $[\mathrm{M}_{\odot}]$ &
        $[\mathrm{dex}]$ \\
        \midrule
\multirow{18}{*}{TNG-Cluster}
        & \multirow{9}{*}{Last host: stripping environment}
        & \multirow{3}{*}{$0.5$--$1.0 \times 10^{14}\,\mathrm{M}_{\odot}$}
        & Main halo
        & 20
        & $10.81^{+0.06}_{-0.06}$
        & $11.16^{+0.14}_{-0.05}$
        & $0.26$ \\

        & 
        & 
        & Infall in a group
        & 11
        & $10.20^{+0.11}_{-0.10}$
        & $10.39^{+0.14}_{-0.35}$
        & $0.36$ \\

        & 
        & 
        & Infall outside group
        & 20
        & $9.53^{+0.07}_{-0.07}$
        & $9.68^{+0.14}_{-0.15}$
        & $0.33$ \\
        \addlinespace[3pt]

        & 
        & \multirow{3}{*}{$1.0$--$1.5 \times 10^{14}\,\mathrm{M}_{\odot}$}
        & Main halo
        & 20
        & $10.90^{+0.05}_{-0.05}$
        & $11.12^{+0.14}_{-0.08}$
        & $0.22$ \\

        & 
        & 
        & Infall in a group
        & 14
        & $10.22^{+0.12}_{-0.12}$
        & $10.25^{+0.31}_{-0.22}$
        & $0.48$ \\

        & 
        & 
        & Infall outside group
        & 20
        & $9.61^{+0.05}_{-0.06}$
        & $9.73^{+0.07}_{-0.13}$
        & $0.26$ \\
        \addlinespace[3pt]

        & 
        & \multirow{3}{*}{$1.5$--$2.0 \times 10^{14}\,\mathrm{M}_{\odot}$}
        & Main halo
        & 6
        & $11.12^{+0.11}_{-0.10}$
        & $11.33^{+0.24}_{-0.11}$
        & $0.28$ \\

        & 
        & 
        & Infall in a group
        & 6
        & $9.69^{+0.24}_{-0.24}$
        & $9.87^{+0.17}_{-0.39}$
        & $0.65$ \\

        & 
        & 
        & Infall outside group
        & 6
        & $9.59^{+0.07}_{-0.07}$
        & $9.75^{+0.12}_{-0.15}$
        & $0.19$ \\
        \addlinespace[5pt]

        & \multirow{9}{*}{First host: creation environment}
        & \multirow{3}{*}{$0.5$--$1.0 \times 10^{14}\,\mathrm{M}_{\odot}$}
        & Main halo
        & 20
        & $10.49^{+0.05}_{-0.05}$
        & $10.81^{+0.08}_{-0.16}$
        & $0.21$ \\

        & 
        & 
        & Infall in a group
        & 11
        & $9.98^{+0.05}_{-0.05}$
        & $10.01^{+0.17}_{-0.06}$
        & $0.17$ \\

        & 
        & 
        & Infall in a galaxy
        & 20
        & $9.63^{+0.06}_{-0.06}$
        & $9.73^{+0.16}_{-0.11}$
        & $0.28$ \\
        \addlinespace[3pt]

        & 
        & \multirow{3}{*}{$1.0$--$1.5 \times 10^{14}\,\mathrm{M}_{\odot}$}
        & Main halo
        & 20
        & $10.60^{+0.06}_{-0.06}$
        & $10.78^{+0.20}_{-0.08}$
        & $0.26$ \\

        & 
        & 
        & Infall in a group
        & 14
        & $10.12^{+0.05}_{-0.05}$
        & $10.11^{+0.23}_{-0.10}$
        & $0.21$ \\

        & 
        & 
        & Infall in a galaxy
        & 20
        & $9.73^{+0.05}_{-0.05}$
        & $9.87^{+0.06}_{-0.04}$
        & $0.24$ \\
        \addlinespace[3pt]

        & 
        & \multirow{3}{*}{$1.5$--$2.0 \times 10^{14}\,\mathrm{M}_{\odot}$}
        & Main halo
        & 6
        & $10.90^{+0.06}_{-0.07}$
        & $11.24^{+0.06}_{-0.08}$
        & $0.18$ \\

        & 
        & 
        & Infall in a group
        & 6
        & $9.99^{+0.21}_{-0.21}$
        & $10.34^{+0.33}_{-0.22}$
        & $0.54$ \\

        & 
        & 
        & Infall in a galaxy
        & 6
        & $9.83^{+0.08}_{-0.08}$
        & $10.07^{+0.07}_{-0.12}$
        & $0.20$ \\
        \midrule
\multirow{18}{*}{Manhattan}
        & \multirow{9}{*}{Last host: stripping environment}
        & \multirow{3}{*}{$0.5$--$1.0 \times 10^{14}\,\mathrm{M}_{\odot}$}
        & Main halo
        & 3
        & $10.61^{+0.27}_{-0.27}$
        & $10.82^{+0.97}_{-0.75}$
        & $0.59$ \\

        & 
        & 
        & Infall in a group
        & 2
        & $9.80^{+0.69}_{-0.69}$
        & $9.66^{+0.83}_{-0.83}$
        & $0.97$ \\

        & 
        & 
        & Infall outside group
        & 3
        & $10.39^{+0.21}_{-0.21}$
        & $10.82^{+0.07}_{-0.96}$
        & $0.48$ \\
        \addlinespace[3pt]

        & 
        & \multirow{3}{*}{$1.0$--$1.5 \times 10^{14}\,\mathrm{M}_{\odot}$}
        & Main halo
        & 20
        & $11.08^{+0.05}_{-0.05}$
        & $11.32^{+0.21}_{-0.09}$
        & $0.22$ \\

        & 
        & 
        & Infall in a group
        & 16
        & $9.98^{+0.12}_{-0.12}$
        & $9.96^{+0.46}_{-0.39}$
        & $0.49$ \\

        & 
        & 
        & Infall outside group
        & 20
        & $10.23^{+0.06}_{-0.06}$
        & $10.39^{+0.10}_{-0.07}$
        & $0.26$ \\
        \addlinespace[3pt]

        & 
        & \multirow{3}{*}{$1.5$--$2.0 \times 10^{14}\,\mathrm{M}_{\odot}$}
        & Main halo
        & 17
        & $11.10^{+0.06}_{-0.06}$
        & $11.37^{+0.15}_{-0.06}$
        & $0.25$ \\

        & 
        & 
        & Infall in a group
        & 13
        & $10.15^{+0.17}_{-0.17}$
        & $10.35^{+0.47}_{-0.75}$
        & $0.64$ \\

        & 
        & 
        & Infall outside group
        & 17
        & $10.33^{+0.08}_{-0.08}$
        & $10.44^{+0.20}_{-0.05}$
        & $0.34$ \\
        \addlinespace[5pt]

        & \multirow{9}{*}{First host: creation environment}
        & \multirow{3}{*}{$0.5$--$1.0 \times 10^{14}\,\mathrm{M}_{\odot}$}
        & Main halo
        & 3
        & $10.70^{+0.12}_{-0.12}$
        & $10.76^{+0.45}_{-0.24}$
        & $0.23$ \\

        & 
        & 
        & Infall in a group
        & 3
        & $10.23^{+0.26}_{-0.26}$
        & $10.27^{+0.49}_{-0.19}$
        & $0.46$ \\

        & 
        & 
        & Infall in a galaxy
        & 3
        & $10.08^{+0.16}_{-0.16}$
        & $10.05^{+0.47}_{-0.28}$
        & $0.35$ \\
        \addlinespace[3pt]

        & 
        & \multirow{3}{*}{$1.0$--$1.5 \times 10^{14}\,\mathrm{M}_{\odot}$}
        & Main halo
        & 20
        & $10.80^{+0.04}_{-0.04}$
        & $10.88^{+0.08}_{-0.04}$
        & $0.17$ \\

        & 
        & 
        & Infall in a group
        & 16
        & $10.18^{+0.06}_{-0.06}$
        & $10.27^{+0.14}_{-0.13}$
        & $0.25$ \\

        & 
        & 
        & Infall in a galaxy
        & 20
        & $10.39^{+0.05}_{-0.05}$
        & $10.59^{+0.04}_{-0.18}$
        & $0.24$ \\
        \addlinespace[3pt]

        & 
        & \multirow{3}{*}{$1.5$--$2.0 \times 10^{14}\,\mathrm{M}_{\odot}$}
        & Main halo
        & 17
        & $10.91^{+0.04}_{-0.03}$
        & $11.07^{+0.02}_{-0.10}$
        & $0.14$ \\

        & 
        & 
        & Infall in a group
        & 13
        & $10.18^{+0.06}_{-0.06}$
        & $10.36^{+0.17}_{-0.08}$
        & $0.23$ \\

        & 
        & 
        & Infall in a galaxy
        & 17
        & $10.36^{+0.05}_{-0.05}$
        & $10.48^{+0.17}_{-0.09}$
        & $0.21$ \\
        \bottomrule
    \end{tabular}%
    }
    \vspace{0.5cm}
    \caption{
    Host stellar mass statistics for TNG-Cluster and Manhattan. The table reports the mean and median stellar masses of the identified host galaxies, expressed as $\rm \log_{10}(M_{\star,\mathrm{host}}/\mathrm{M}_{\odot})$, for each host channel and protocluster mass range at $z\simeq2$. The first block classifies hosts by the stripping environment of the last identified host. The second block classifies the first identified host by the creation environment of the ICL particle. Only ICL particles with an identified host satisfying $\rm M_{\star,\mathrm{host}}\geq10^7\,\mathrm{M}_{\odot}$ are included. For each protocluster and channel, host masses are weighted by the sampled ICL particle mass. Quoted uncertainties correspond to the $16$th--$84$th percentile intervals from bootstrap resampling of protoclusters. The final column gives the protocluster-to-protocluster scatter in the per-protocluster mean host mass.
    }
    \label{tab:hosts}
\end{table*}

\clearpage

\section{\centering Impact of galaxy definition}
\label{B}

We tested how our galaxy size definition would affect our ICL mass results. To do that, we estimate the amount of ICL in the main halo and infall region for the full sample of protoclusters analysed in this work at $z=2$ using different values of $\rm \times r_{half}$. Note that the BCG aperture is also variable as the other satellite galaxies.

\begin{figure}
\centering
\vspace{-0.3cm}
\includegraphics[width=0.4\columnwidth]{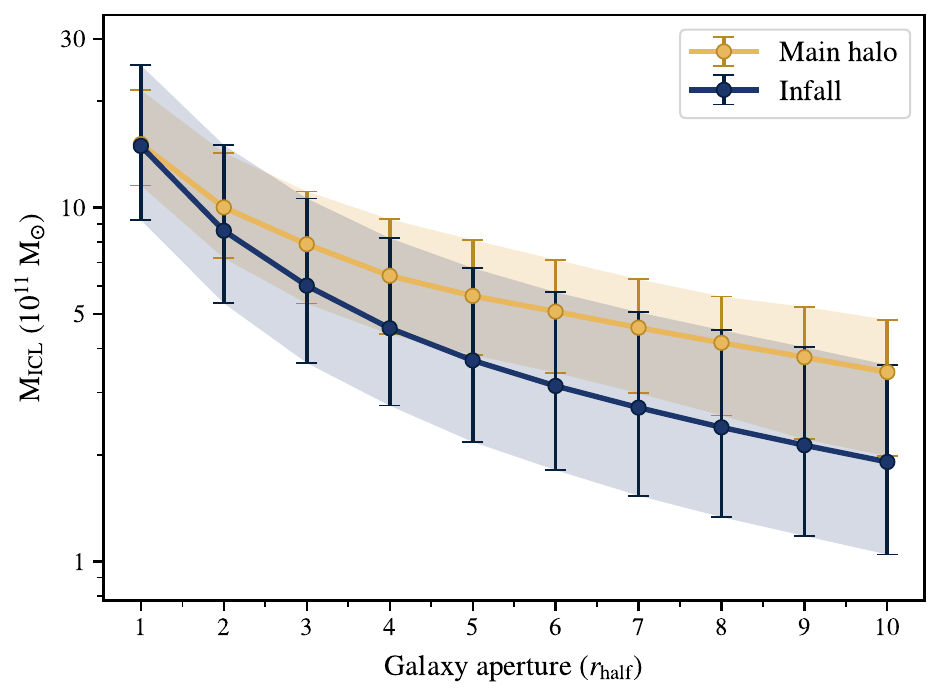}
\vspace{-0.3cm}
\caption{Stellar mass of the ICL at redshift 2 in the main halo and infall regions as a function of galaxy aperture, in units of the stellar half-mass radius for the full sample of protoclusters in TNG-Cluster. The ICL stellar mass decreases as the galaxy aperture increases, but for reasonable values supported by the literature (1-4 $\rm r_{half}$), we still find a considerable amount of ICL in both environments. Note that the BCG aperture is also variable as the satellite galaxies, so for 2x$\rm r_{half}$ the BCG aperture is also 2x$\rm r_{half}$, not 3x$\rm r_{half}$ as in the main analysis.}
\label{galdef}
\vspace{-0.53cm}
\end{figure}

As expected, the inferred ICL mass decreases as the adopted galaxy aperture increases, since more stellar particles are assigned to galaxies rather than to the diffuse component. However, we still find a substantial amount of ICL in both the main halo and the infall region for apertures up to 4$\rm \times r_{half}$, which is approximately the largest galaxy aperture supported by previous observational studies \citep{Werner_2023}. Fig.~\ref{galdef} shows how the inferred ICL mass varies as a function of the galaxy aperture. The shaded regions show the 16th-84th percentiles considering all protoclusters in the TNG-Cluster sample used in this work.

Fig.~\ref{galdef} also shows that there is a reasonable amount of ICL in the main halo and infall components for all aperture choices. Although the absolute ICL mass depends on the aperture choice, the qualitative interpretation of substantial ICL in the main halo and the infall is preserved for all apertures. We conclude that the qualitative results presented in this paper are robust.

\begin{figure}
\centering
\includegraphics[width=1\textwidth]{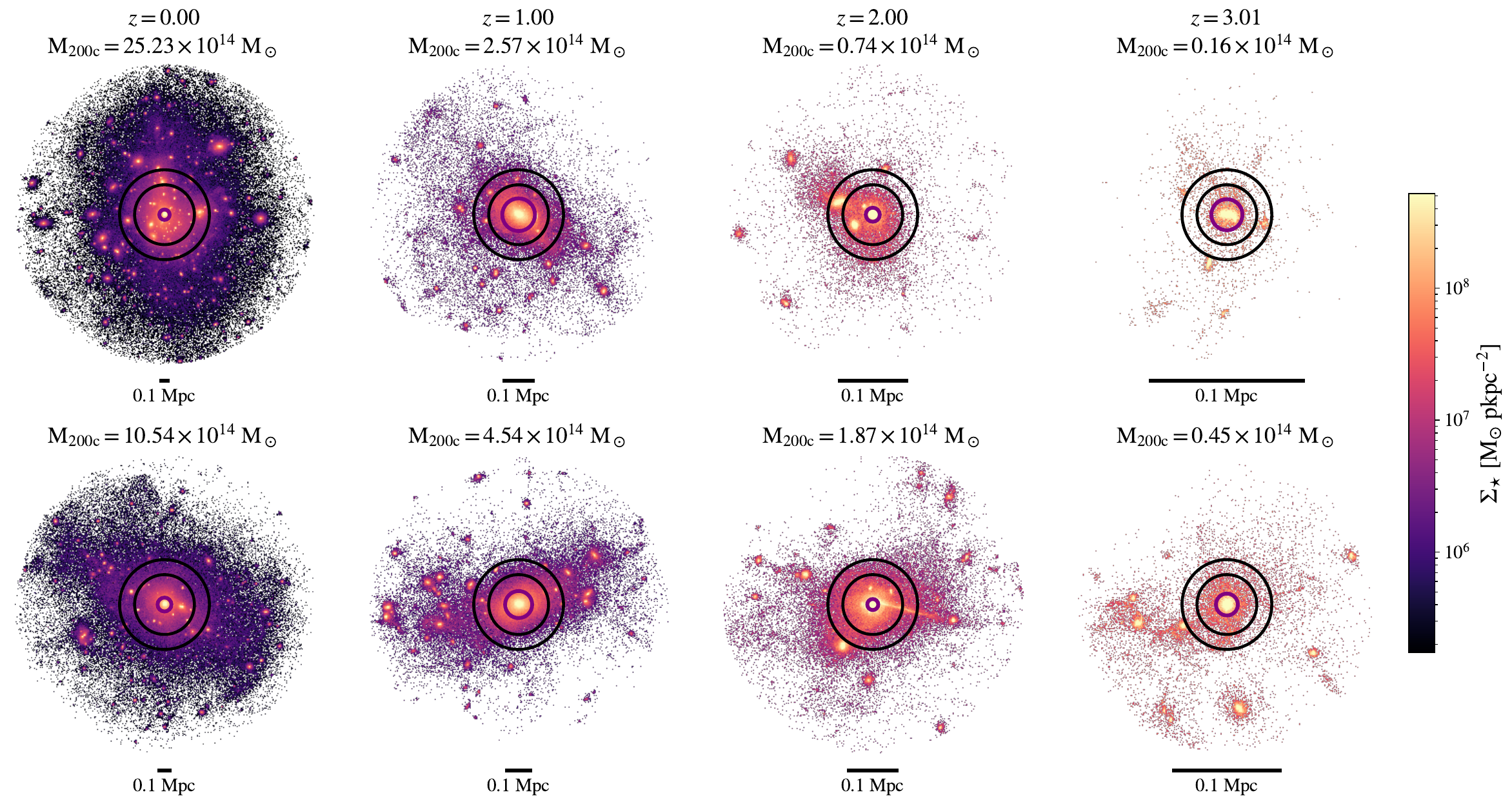}
\vspace{-0.7cm}
\caption{The evolution of the stellar component in $\rm 0.5 \times R_{200c}$ of the most massive protocluster in TNG-Cluster (bottom panel) and the most massive cluster (upper panel). The inner purple circle represents $\rm 3\times r_{half}$ of the central galaxy and the outer black circles represent $\rm 0.1-0.15 \times R_{200c}$. The different panels show the stellar distributions at different redshifts, and $\rm M_{200c}$ is also given, 0.1 Mpc scale bars can be seen in the bottom.}
\label{ICL_tng2}
\vspace{-0.5cm}
\end{figure}

Additionally, we attach an image of the (proto)cluster main halo in TNG-Cluster showing the inner apertures of $\rm 0.1-0.15 \times R_{200c}$ and the central galaxy aperture adopted (Fig.\,\ref{ICL_tng2}). Although for low redshift studies the aperture of 2$\rm \times r_{half}$ is adopted, we find that this value might be too small for central galaxies at $z>2$, so we decide to be conservative in our ICL estimates and adopt an aperture of  3$\rm \times r_{half}$ only for the central galaxy.

\label{lastpage}

\end{document}